\documentclass[pdflatex,11pt]{article}

\usepackage{graphicx}
\usepackage{amsmath,amssymb}
\usepackage{hyperref}
\usepackage{geometry}
\newcommand{\diff}{\mathrm{d}}
\newcommand{\dmeasure}[2]{\mathrm{d}^{#1} #2 \,}

\newcommand{\tr}{\mathrm{tr}}

\newcommand{\psibar}{\bar{\psi}}
\graphicspath{{graph/}}
\begin{document}

\title{Four-Fermion Interaction Model on $\mathcal{M}^{D-1} \otimes S^1$}
\author{Tomohiro Inagaki ${}^{1, 2} \footnote{\href{mailto:inagaki@hiroshima-u.ac.jp}{inagaki@hiroshima-u.ac.jp}}$ ,
 Yamato Matsuo ${}^{3} \footnote{\href{mailto:ya-matsuo@hiroshima-u.ac.jp}{ya-matsuo@hiroshima-u.ac.jp}}$ ,
 and Hiromu Shimoji ${}^{3} \footnote{\href{mailto:h-shimoji@hiroshima-u.ac.jp}{h-shimoji@hiroshima-u.ac.jp}}$}
\date{\it
${}^1$Information Media Center, Hiroshima University, Higashi-Hiroshima, 739-8521, Japan,\\
${}^2$Core of Research for the Energetic Universe, Hiroshima University, Higashi-Hiroshima, 739-8526, Japan,\\
${}^3$Department of Physics, Hiroshima University, Higashi-Hiroshima, 739-8526, Japan}

\maketitle
\begin{abstract}
 Four-fermion interaction models are often used as simplified models of interacting fermion fields with the chiral symmetry. The chiral symmetry is dynamically broken for a larger four-fermion coupling. It is expected that the broken symmetry is restored under extreme conditions. In this paper, the finite size effect on the chiral symmetry breaking is investigated in the four-fermion interaction model. We consider the model on a flat spacetime with a compactified spatial coordinate, $\mathcal{M}^{D-1} \otimes S^1$ and obtain explicit expressions of the effective potential for arbitrary spacetime dimensions in the leading order of the $1/N$ expansion. Evaluating the effective potential, we show the critical lines which divide the symmetric and the broken phase and the sign-flip condition for the Casimir force.
\end{abstract}
\section{Introduction}

Fundamental theories of particle physics are constructed based on several types of symmetry. It is expected that a fundamental theory with a higher symmetry is realized at the early universe. The symmetry of the theory is partly broken on the ground state. The remnant symmetry is observed in our laboratories. Since the physical state depends on the environment, the broken symmetry can be restored under extreme conditions, small size, high temperature, high density, strong curvature, strong electromagnetic field, and so on. It is considered that there is a possibility to test the models of particle physics through the critical phenomena induced by the symmetry transition with changing the environment.

Y.~Nambu and G.~Jona-Lasinio proposed a simple model with interacting fermions in 1961 \cite{Nambu:NJL}. A four-fermion interaction model  with a discrete $Z_2$ chiral symmetry is introduced in two dimensions by D.~J.~Gross and A.~Neveu \cite{Gross:1974jv}. In the Nambu--Jona-Lasinio (NJL) and Gross--Neveu (GN) models four-fermion interactions induce non-vanishing expectation value for the composite operator constructed by a fermion and an anti-fermion and the chiral symmetry is spontaneously broken. Many works have been done to study the symmetry transition in the four-fermion interaction models under various environmental conditions, for a review, see  \cite{Rosenstein:1990nm, Klevansky:1992qe,Hatsuda:1994pi,Inagaki:1997kz,Miransky:2015ava} and references therein. One of the interesting conditions to induce the symmetry transition is found in the size of the spacetime. All the materials may be confined inside a small size space with a non-trivial topology at the early universe. The existence of finite extra dimensions is predicted in string theory and M-theory.

The finite size effect on the chiral symmetry has been studied in four-fermion interaction models  with periodic or anti-periodic boundary conditions for the fermion fields. It is found that the chiral symmetry tends to be broken due to the finite size effect for the periodic boundary condition and the broken symmetry tends to be restored for the anti-periodic boundary condition \cite{Inagaki:1997kz, Kim:1994es, Ishikawa:1998uu}. The effective potential of the four-fermion interaction model  has been calculated in a more general $\mathrm{U}(1)$-valued boundary condition along with a compact direction and the phase structure of the model is evaluated with respect to the $\mathrm{U}(1)$ phase \cite{Flachi:2013}.

In an Abelian gauge theory such boundary conditions can be realized for charged fermions through the Aharonov-Bohm effect \cite{Aharonov:1959fk}. Hence, the finite size effect with a $\mathrm{U}(1)$-valued boundary condition can be realized in the presence of a gauge field. It is well-known that a constant magnetic field enhances the chiral symmetry breaking. It is shown that the enhanced symmetry breaking can be counteracted by the finite size effect with the anti-periodic boundary condition \cite{Ferrer:1999gs}. A constant magnetic flux crosses the transverse section of the cylinder has been studied in lower dimensional cylindrical spacetime \cite{Gamayun:2004wv, Ebert:2015, Zhokhov:2016xzp, Chernodub:2019nct}. The possibility of an inhomogeneous condensation has been discussed in a superconducting ring with an Aharonov-Bohm magnetic flux \cite{Yoshii:2014fwa}. It has been pointed out that the finite size phase transition may be observed as a nontrivial behavior in the Casimir force \cite{Casimir:1948dh, Bordag:1995gm, Bordag:2001qi}.

In the present paper we study four-fermion interaction models on $\mathcal{M}^{D-1} \otimes S^1$ and develop the procedure to calculate the stable environment and the Casimir effect. The ground state is found by observing the minimum of the effective potential. The phase boundary can be found by solving the stationary condition of the effective potential for a fixed size and a topology. We need to evaluate the zero-point energy to determine the stable size and the topology and to calculate the Casimir force.

The paper is organized as follows. In Sec.~2 we briefly review the chiral symmetry breaking in four-fermion interaction models on $\mathcal{M}^{D}$. We employ the $1/N$ expansion and calculate the effective potential for the fermion and anti-fermion composite field. In Sec.~3 we consider the model on $\mathcal{M}^{D-1} \otimes S^1$ with a nontrivial topology. Evaluating the effective potential with a zero-point energy, we obtain the dynamically generated fermion mass and the critical length for a fixed boundary condition. In Sec.~4 the Casimir force is calculated from the effective potential and the sign flip condition is derived. Section 5 is devoted to the concluding remarks.

 \section{The basic model: Four-fermion interaction model on $\mathcal{M}^D$}
 
 In this section we consider the Dirac fermion on a flat $D$-dimensional Minkowski spacetime $\mathcal{M}^D$ and follow the discussions in \cite{Inagaki:1994ec}.
The Dirac fermion is decomposed into left- and right-handed chiral states. In relativistic quantum field theories left- and right-handed chiral state, $\psi_L$ and $\psi_R$, for the four-components Dirac fermion $\psi$ can be described as
\begin{align}
 \psi_L =\frac{1-\gamma^5}{2}\psi ,\hspace{2ex} \psi_R =\frac{1+\gamma^5}{2}\psi ,
\label{left:right}
\end{align}
with the fifth Dirac gamma matrix, $\gamma^5$. The chiral symmetry which preserves the chirality of the system gives fundamental and important concepts in particle physics. It is defined by the invariance under the chiral transformation,
\begin{align}
 \psi \longrightarrow e^{i\gamma^5\theta}\psi.
\label{trans:chiral}
\end{align}
It transforms the left-handed and right-handed fermions with an opposite sign phase. The chiral symmetry prohibits the fermion field from having a mass term. The simplest fermion and anti-fermion interaction which maintains the chiral symmetry is four-fermion interactions.

Throughout this paper we employ a simple four-fermion interaction model with $N$-flavor of Dirac fermions which is introduced by Y.~Nambu and G.~Jona-Lasinio \cite{Nambu:NJL}. The model is defined by the action,
 \begin{align}
  S=
  \int \dmeasure{D}{x}\left[ \sum_{a=1}^{N}\psibar_a i \gamma^\mu \partial_\mu \psi_a + \frac{\lambda_0}{2N} \left( \left(\sum_{a=1}^{N}\psibar_a \psi_a\right)^2 + \left( \sum_{a=1}^{N}\psibar_a i \gamma^5 \psi_a \right)^2\right) \right],
  \label{NJL:action}
 \end{align}
 where the index $a$ denotes the flavors of the fermion field $\psi$ and $\lambda_0$ is the bare coupling constant for the four-fermion interactions. The action \eqref{NJL:action} is invariant under the chiral transformation \eqref{trans:chiral}. If the four-fermion interaction induces a non-vanishing expectation value for the composite operator $\psibar_a \psi_a$, a fermion mass term is generated and the chiral symmetry is spontaneously broken.

 For practical calculations it is more convenient to introduce the auxiliary field, $\sigma$ and $\pi$, and rewrite the action as,
\begin{align}
  S=
  \int \dmeasure{D}{x}\left[ \sum_{a=1}^{N}\psibar_a \left( i \gamma^\mu \partial_\mu -\sigma -i \pi \gamma^5 \right) \psi_a - \frac{N}{2\lambda_0} \left( \sigma^2 + \pi^2 \right) \right].
  \label{NJL:action:ax}
 \end{align}
This action describes the same theory with the action \eqref{NJL:action}. The original action \eqref{NJL:action} is reproduced by substituting the solutions of the classical equation of motions
 \begin{align}
  \sigma = -\frac{\lambda_0}{N}\sum_{a=1}^{N}\psibar_a \psi_a,  \hspace{2ex} \pi = -\frac{\lambda_0}{N}\sum_{a=1}^{N}\psibar_a i\gamma^5\psi_a.
 \end{align}

Performing the path integral for the Dirac fermion and assuming homogeneous expectation values for $\sigma$ and $\pi$, we obtain the effective potential at the leading order of the $1/N$ expansion,
 \begin{align}
  V_0(\sigma,\pi) = \frac{1}{2\lambda_0} (\sigma^2+\pi^2) + i \int \frac{\dmeasure{D}{k}}{(2\pi)^D} \tr\ln \frac{ \gamma^\mu k_\mu - \sigma -i\gamma^5\pi}{-\omega}, \label{EFonMD0}
 \end{align}
 where the trace, $\tr$, stands for the sum over the Dirac indices and $\omega$ is an arbitrary mass scale. Due to the chiral symmetry of the action, we set $\pi=0$ without loss of generality.
Then the expectation value of $\sigma$ under the ground state is determined by observing the minimum of the effective potential,
  \begin{align}
  V_0(\sigma) = \frac{1}{2\lambda_0} \sigma^2 + i \int \frac{\dmeasure{D}{k}}{(2\pi)^D} \tr\ln \frac{ \gamma^\mu k_\mu - \sigma}{-\omega}. \label{EFonMD}
 \end{align}
 If the auxiliary field, $\sigma$, develops a non-vanishing expectation value, the fermion acquires a non-vanishing mass and the chiral symmetry is broken. Thus we regard the auxiliary field, $\sigma$, as an order parameter for the chiral symmetry breaking.  It is noted that the effective potential \eqref{EFonMD} coincides with the one in the GN model \cite{Gross:1974jv}. The GN model has the discrete $Z_2$ chiral symmetry under the transformation $\psi \rightarrow \gamma^5 \psi$. Since a continuous symmetry cannot be broken in two dimensions, we employ the GN model and evaluate the discrete chiral symmetry breaking in two dimensions.
 
 We usually shift the origin of the effective potential to zero and remove the divergent zero-point energy,
 \begin{align}
  \tilde{V}_0(\sigma)
  &\equiv V_0(\sigma) -V_0(\sigma =0) \nonumber \\
  &= \frac{1}{2\lambda_0}\sigma^2
   + i\int\frac{\dmeasure{D}{k}}{(2\pi)^D} \tr \ln \left( \frac{\gamma^\mu k_\mu - \sigma}{\gamma^\mu k_\mu} \right) .
   \label{epot0}
 \end{align}
The mass scale, $\omega$, dependence is eliminated by this subtraction.
Integrating over the momentum space, the effective potential reads,
 \begin{align}
  \tilde{V}_0(\sigma)
  =\frac{1}{2\lambda_0}\sigma^2 - \frac{\tr I }{(4\pi)^{D/2} D}\Gamma\left(1-\frac{D}{2}\right)(\sigma^2)^\frac{D}{2} .
 \end{align}
We set $\tr I = 2^{D/2}$ for numerical calculations. The bare four-fermion coupling $\lambda_0$ is replaced with the renormalized one, $\lambda_r$, by imposing the renormalization condition
 \begin{align}
  \left. \frac{\partial^2 \tilde{V}_0(\sigma)}{\partial \sigma^2}\right|_{\sigma = \mu}
  = \frac{1}{\lambda_0}-\frac{\tr I (D-1)}{(4\pi)^{D/2}}\Gamma\left(1-\frac{D}{2}\right)\mu^{D-2}
  \equiv \frac{1}{\lambda_r}\mu^{D-2}, \label{RC}
 \end{align}
 where $\mu$ denotes the renormalization scale. Therefore the renormalized effective potential is given by
 \begin{align}
  \frac{\tilde{V}_0(\sigma)}{\mu^D}
  =\frac{1}{2}\left( \frac{1}{\lambda_r} -\frac{1}{\lambda_c}\right)\left(\frac{\sigma}{\mu}\right)^2 - \frac{\tr I}{(4\pi)^{D/2} D} \Gamma\left(1-\frac{D}{2}\right)\left(\frac{\sigma^2}{\mu^2}\right)^\frac{D}{2}, \label{EFonMDmu}
 \end{align}
 with
 \begin{align}
  \frac{1}{\lambda_c} =- \frac{\tr I (D-1)}{(4\pi)^{D/2}}\Gamma\left(1-\frac{D}{2}\right).
 \end{align}
Since the four-fermion interaction is not renormalizable in four dimensions, the renormalized effective potential is still divergent for $D=4$. We regard the model at $4-\varepsilon$ dimensions as a regularized model in four dimensions.
 
The expectation value of the auxiliary field, $\sigma$ is obtained as the non-trivial solution of the gap equation, a necessary condition for the minimum of the effective potential,
 \begin{align}
  \left. \frac{\partial \tilde{V}_0(\sigma)}{\partial \sigma}\right|_{\sigma=m_0} =0.
 \end{align}
Solving the gap equation, we obtain the expression for the dynamically generated fermion mass,
 \begin{align}
  m_0 = \mu \left[ \frac{ (4\pi)^{D/2}}{\tr I \cdot \Gamma\left(1-\frac{D}{2}\right)} \left(\frac{1}{\lambda_r}-\frac{1}{\lambda_{c}}\right) \right]^{\frac{1}{D-2}} \label{DynamicallyGeneratedMass}.
 \end{align}
If this expression has a real and non-vanishing value, the fermion mass is dynamically generated. To find the critical value of the coupling constant we take the massless limit $m_0 \to 0$ of equation \eqref{DynamicallyGeneratedMass}. Then we get
 \begin{align}
  \lambda_{cr} = \lambda_c .
 \end{align}
\begin{figure}[tbp]
 \begin{center}
  \includegraphics[width=0.5\hsize]{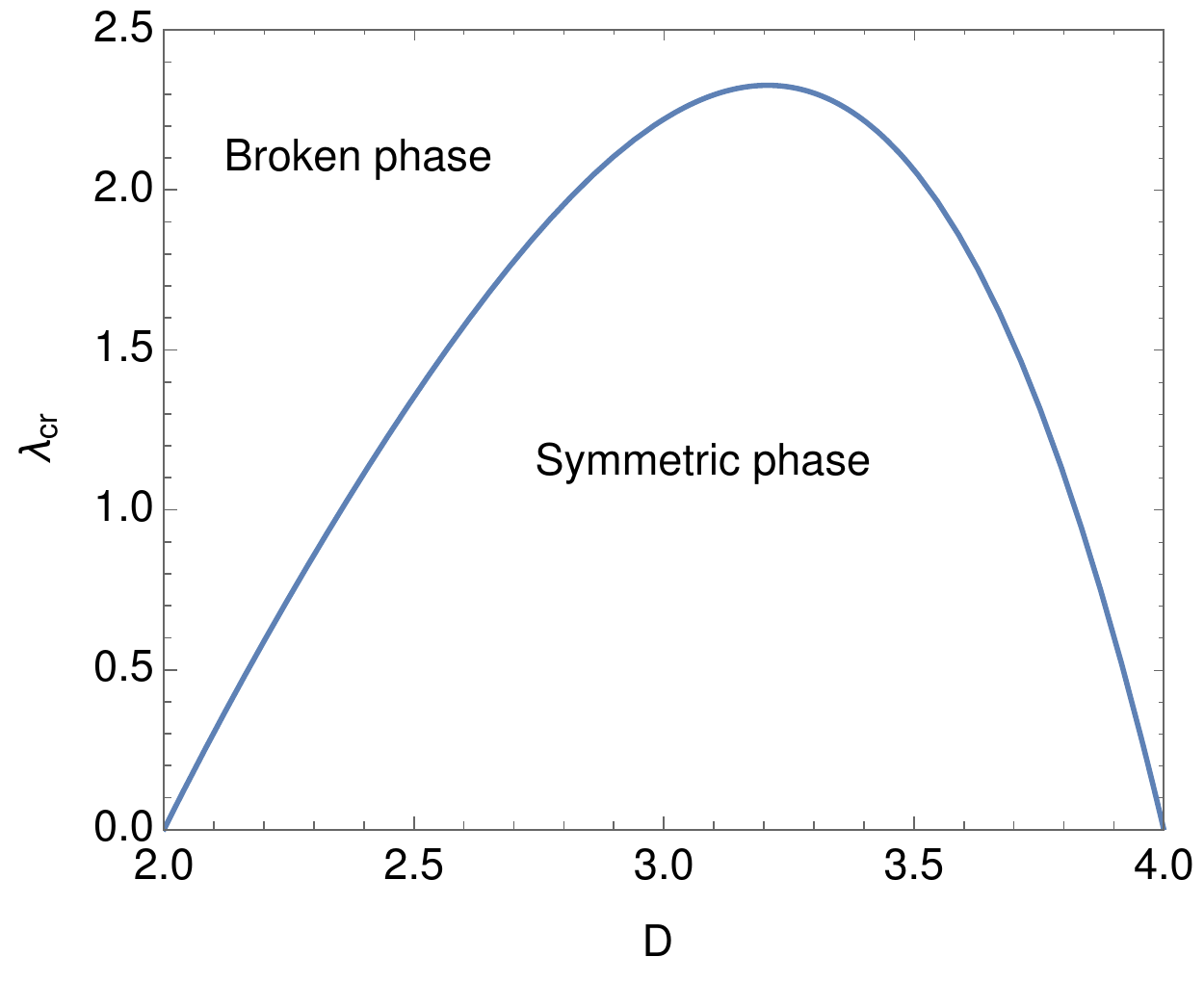}
  \caption{Critical point as the function of the dimension on $\mathcal{M}^D$.}
  \label{criticalpoint}
 \end{center}
\end{figure}
When the four-fermion coupling, $\lambda_r$, is larger than the critical one, $ \lambda_{cr}$, non-vanishing fermion mass is generated and the chiral symmetry is broken. In Fig. \ref{criticalpoint} the critical coupling, $\lambda_{cr}$, is plotted as a function of the spacetime dimension, $D$. Above the line the four-fermion coupling is strong enough to break the chiral symmetry. As is observed in the figure, only the broken phase is realized in two dimensions.

\begin{figure}[htbp]
 \begin{center}
  \includegraphics[width=0.5\hsize]{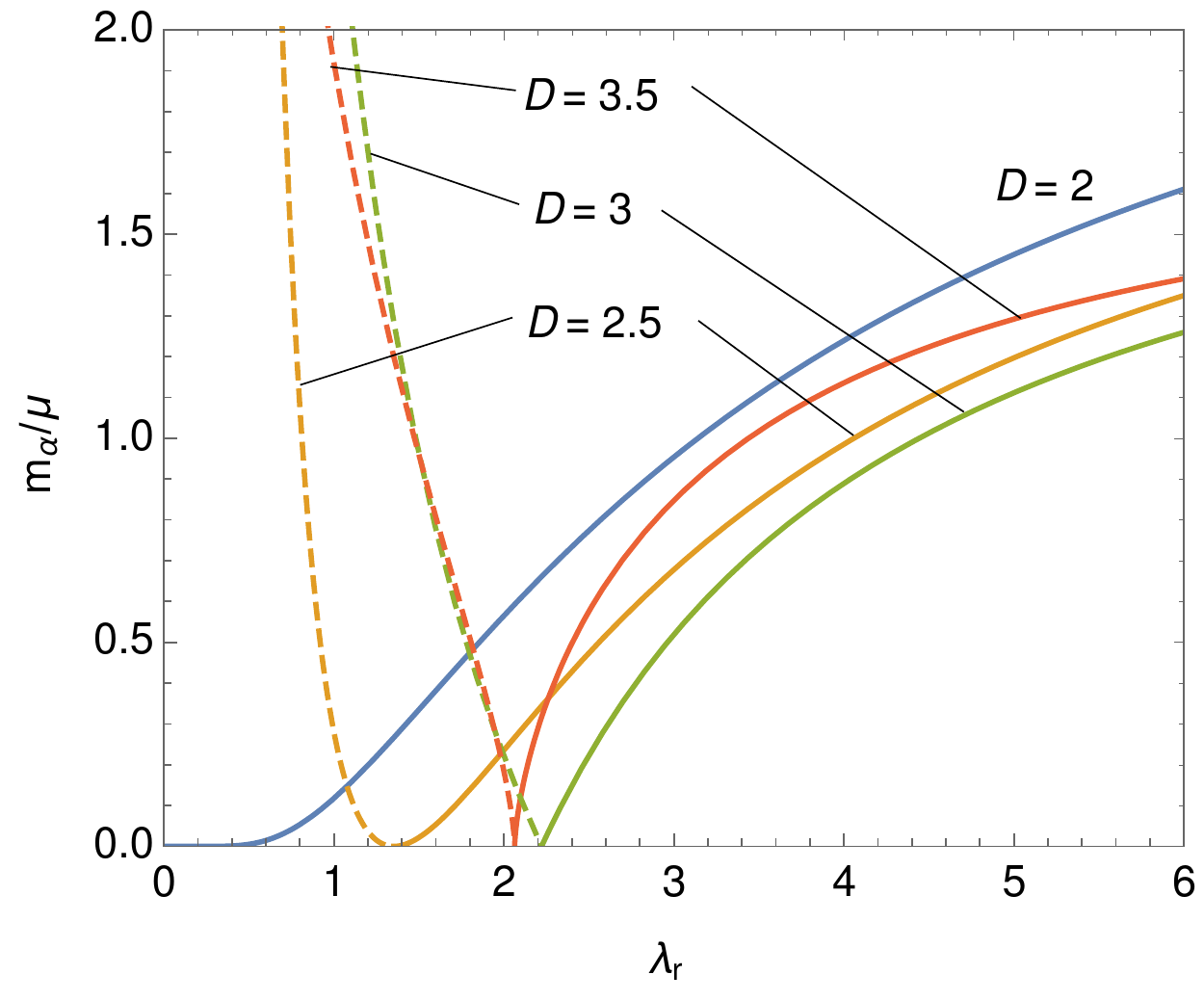}
  \caption{Dynamically generated mass, $m_0$, (solid lines)
  and the mass scale, $m_1$, (dashed lines).
  }
  \label{DGMass}
 \end{center}
\end{figure}

In Fig.~\ref{DGMass} we draw the behavior of the dynamically generated fermion mass, $m_0$, as a function of the renormalized coupling, $\lambda_r$. For $\lambda_r <\lambda_{cr}$ the mass scale, $m_1$, is defined by the absolute value of the right hand side of \eqref{DynamicallyGeneratedMass}. The scale, $m_1$, is also plotted in the figure. Using the dynamically generated mass, $m_0$, and the mass scale, $m_1$, the expression of the effective potential \eqref{EFonMDmu} is simplifies to
\begin{align}
 \frac{\tilde{V}_0(\sigma)}{{m_{0}}^D}
 =\frac{\tr I}{(4\pi)^{D/2}} \Gamma\left(1-\frac{D}{2}\right)\left[\frac{1}{2}\left(\frac{\sigma}{m_0}\right)^2 - \frac{1}{D}\left(\frac{\sigma^2}{m_0^2}\right)^\frac{D}{2}\right], \mbox{ for } \lambda_r > \lambda_{cr}, \label{EFonMDmu:sc}
\end{align}
and
\begin{align}
 \frac{\tilde{V}_0(\sigma)}{{m_{1}}^D}
 =\frac{\tr I}{(4\pi)^{D/2}} \Gamma\left(1-\frac{D}{2}\right)\left[-\frac{1}{2}\left(\frac{\sigma}{m_1}\right)^2 - \frac{1}{D}\left(\frac{\sigma^2}{m_1^2}\right)^\frac{D}{2}\right], \mbox{ for } \lambda_r < \lambda_{cr}, \label{EFonMDmu:wc}
\end{align}
respectively. Since the renormalization scale, $\mu$, and the renormalized coupling, $\lambda_r$, are not independent at the leading order of the $1/N$ expansion, the parameters, $\mu$, and, $\lambda_r$, are rewritten by $m_0$, and $m_1$.  We discuss the phase structure of the four-fermion interaction model, starting from the effective potential \eqref{EFonMDmu:sc} and \eqref{EFonMDmu:wc}. Since the critical coupling is vanishing, $\lambda_{cr}=0$, in two dimensions, only the expression \eqref{EFonMDmu:sc} is adopted for $D=2$.

\section{Four-fermion interaction model on $\mathcal{M}^{D-1} \otimes S^1$}

To study the finite size effect on the dynamical chiral symmetry breaking we consider the four-fermion interaction model \eqref{NJL:action} on a flat spacetime, $\mathcal{M}^{D-1}\otimes S^1$, with one spatial compact direction. The Dirac field on the space is constrained by the size of the compactified space and the boundary condition. We assign the following boundary condition in the compactified direction, $x^{D-1}$,
\begin{align}
 \psi(x^0, \dots, x^{D-1}+L) &= e^{i\pi\delta} \psi(x^0, \dots, x^{D-1}),
 \label{bc}
\end{align}
where $L$ is the length of the compactified space and $\delta$ is the phase factor. For the spatial compact direction the phase factor, $\delta$, is a parameter which is fixed by an environment outside of the system or a non-trivial topology of the very early universe. According to the standard procedure on a compact spacetime, Green functions on $\mathcal{M}^{D-1}\otimes S^1$ are given from the one on $\mathcal{M}^{D}$ by the replacements
\begin{align}
 \left\{
 \begin{aligned}
  \int \frac{\dmeasure{}{k^{D-1}}}{2\pi} &\to \frac{1}{L} \sum_{n=-\infty}^{\infty}, \\
  k^{D-1} &\to \omega_{n} = \frac{2\pi}{L}\left(n + \frac{\delta}{2} \right).
 \end{aligned}
 \right.
\end{align}
The boundary condition \eqref{bc} is satisfied by introducing the discrete variable, $\omega_n$.

Applying these replacements on \eqref{EFonMD}, we obtain the effective potential on $\mathcal{M}^{D}$ at the leading order of the $1/N$ expansion,
\begin{align}
 V(\sigma)
 = \frac{1}{2\lambda_0}\sigma^2 +  \frac{i}{L}\sum_{n=-\infty}^{\infty} \int\frac{\dmeasure{D-1}{K}}{(2\pi)^{D-1}} \tr\ln \frac{ \gamma^\mu K_\mu - \omega_n \gamma^{D-1}- \sigma }{-\omega} , \label{EFwithFinite}
\end{align}
where $K^\mu$ denotes $\{k^0,\cdots,k^{D-2} \}$, the momentum on $\mathcal{M}^{D-1}$. The divergent zero-point energy is removed by the shift of the effective potential,
\begin{align}
 \tilde{V}(\sigma) \equiv V(\sigma) -V_0(\sigma =0) .
\end{align}
To compare the potential energy between the different size, $L$, and phase, $\delta$ we choose the subtracted $V_0(\sigma =0)$ independent of the parameters $L$, $\delta$ and $\sigma$. As is showned in the Appendix, the effective potential, $\tilde{V}(\sigma)$, is represented in several forms. Since the expression which contains no divergent function, $C(L)$, is more convenient for the numerical analysis, we adopt the expression \eqref{Appendix:epot:int}. Thus the normalized effective potential reads,
\begin{align}
 \frac{\tilde{V}(\sigma)}{\mu^D} =& \frac{\tilde{V}_0(\sigma)}{\mu^D} \nonumber \\
 &- \frac{\tr I}{\left(2\sqrt{\pi}\right)^{D-1} \Gamma\left(\frac{D-1}{2}\right)} \frac{1}{L\mu}\int_0^\infty \frac{\dmeasure{}{K}}{\mu} \left(\frac{K}{\mu}\right)^{D-2} \ln \left(2\frac{\cosh\left(L\sqrt{K^2+\sigma^2} \right) -\cos\left( \pi\delta \right)}{\exp\left( L\sqrt{K^2+\sigma^2} \right)} \right). \label{EFwithFiniteSumInt}
\end{align}
The second term in the right hand side describes the finite size corrections with the boundary condition. The term is finite and vanishes at  the $L\to\infty$ limit.

The effective potential \eqref{EFwithFiniteSumInt} depends on the coupling constant, $\lambda_r$, and the renormalization scale, $\mu$. Substituting equations \eqref{EFonMDmu:sc} and \eqref{EFonMDmu:wc} into \eqref{EFwithFiniteSumInt}, the parameters $\lambda_r$ and $\mu$ are described by $m_0$ and $m_1$.
Thus the effective potential normalized reads
\begin{align}
 \frac{\tilde{V}(\sigma)}{m_\alpha^D}
 =& \frac{\tilde{V}_0(\sigma)}{m^D_\alpha}
 \nonumber \\
 &-\frac{\tr I}{ (2\sqrt{\pi})^{D-1}\Gamma\left(\frac{D-1}{2}\right)} \frac{1}{Lm_\alpha} \int_0^\infty \frac{\dmeasure{}{K}}{m_\alpha} \left(\frac{K}{m_\alpha}\right)^{D-2}
 \ln \left(2\frac{\cosh\left(L\sqrt{K^2+\sigma^2} \right) -\cos\left(\pi\delta \right)}{\exp \left(L\sqrt{K^2+\sigma^2} \right)} \right) ,\label{EFSumInt}
\end{align}
where we set $\alpha=0$ and $1$ for $\lambda_r > \lambda_{cr}$ and $\lambda_r < \lambda_{cr}$, respectively.  The reason for this replacement is to reduce the number of parameters. We split the expression with respect to the critical value of the coupling constant, $\lambda_{cr}$, which is defined  on $\mathcal{M}^D$. Though the model on $\mathcal{M}^D$ is different from the one on $\mathcal{M}^{D-1}\otimes S^1$ as long as the length $L$ is finite, the normalization by $m_0 (m_1)$ gives one criterion in considering the compactified model.

\begin{figure}[htbp]
 \begin{center}
  \begin{tabular}{cc}
   \begin{minipage}{0.35\hsize}
    \begin{center}
     \includegraphics[width=0.95\hsize,clip]
     {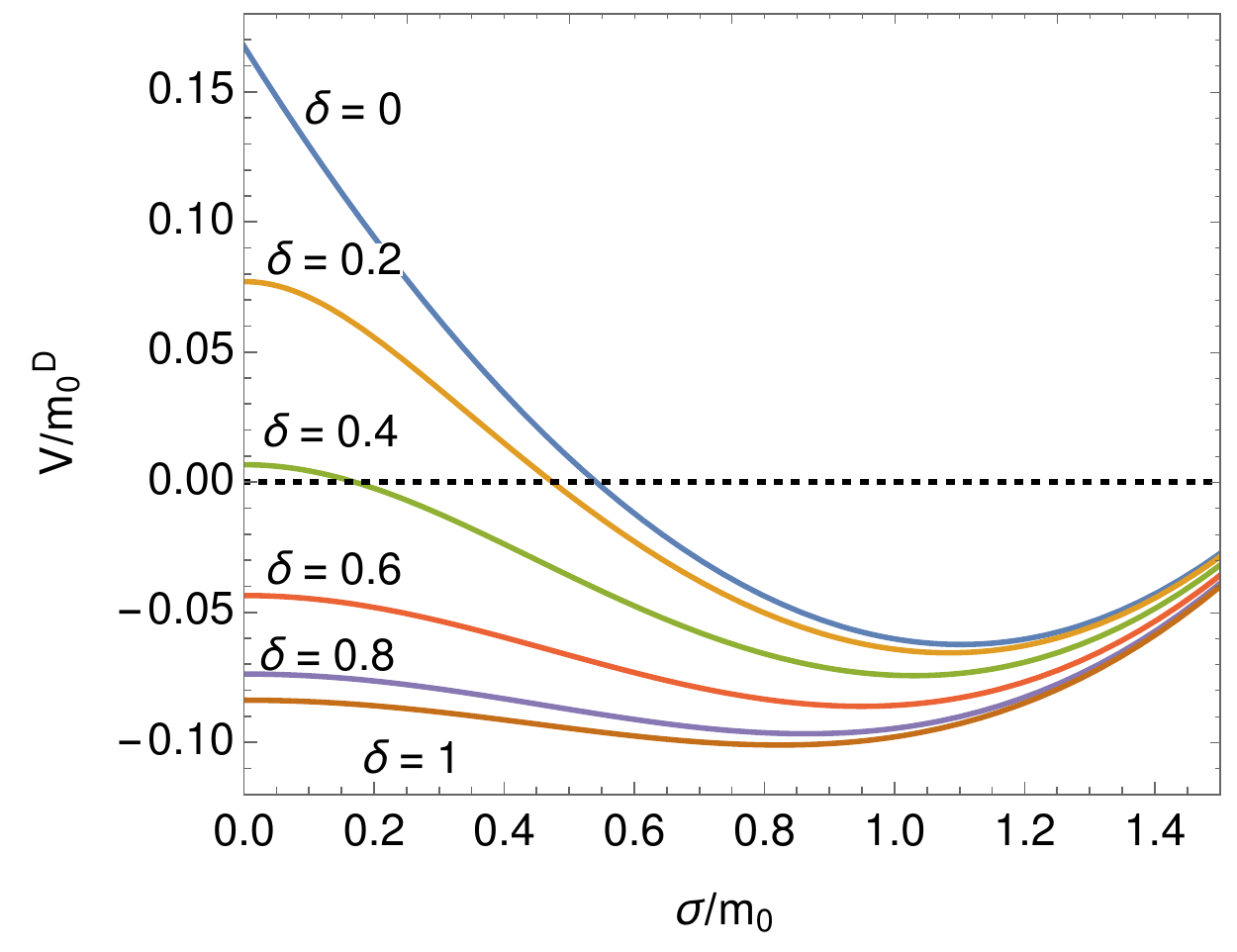}
     (a) $D=2$
    \end{center}
    \vglue 1mm
   \end{minipage}
&
   \begin{minipage}{0.35\hsize}
    \begin{center}
     \includegraphics[width=0.95\hsize,clip]
     {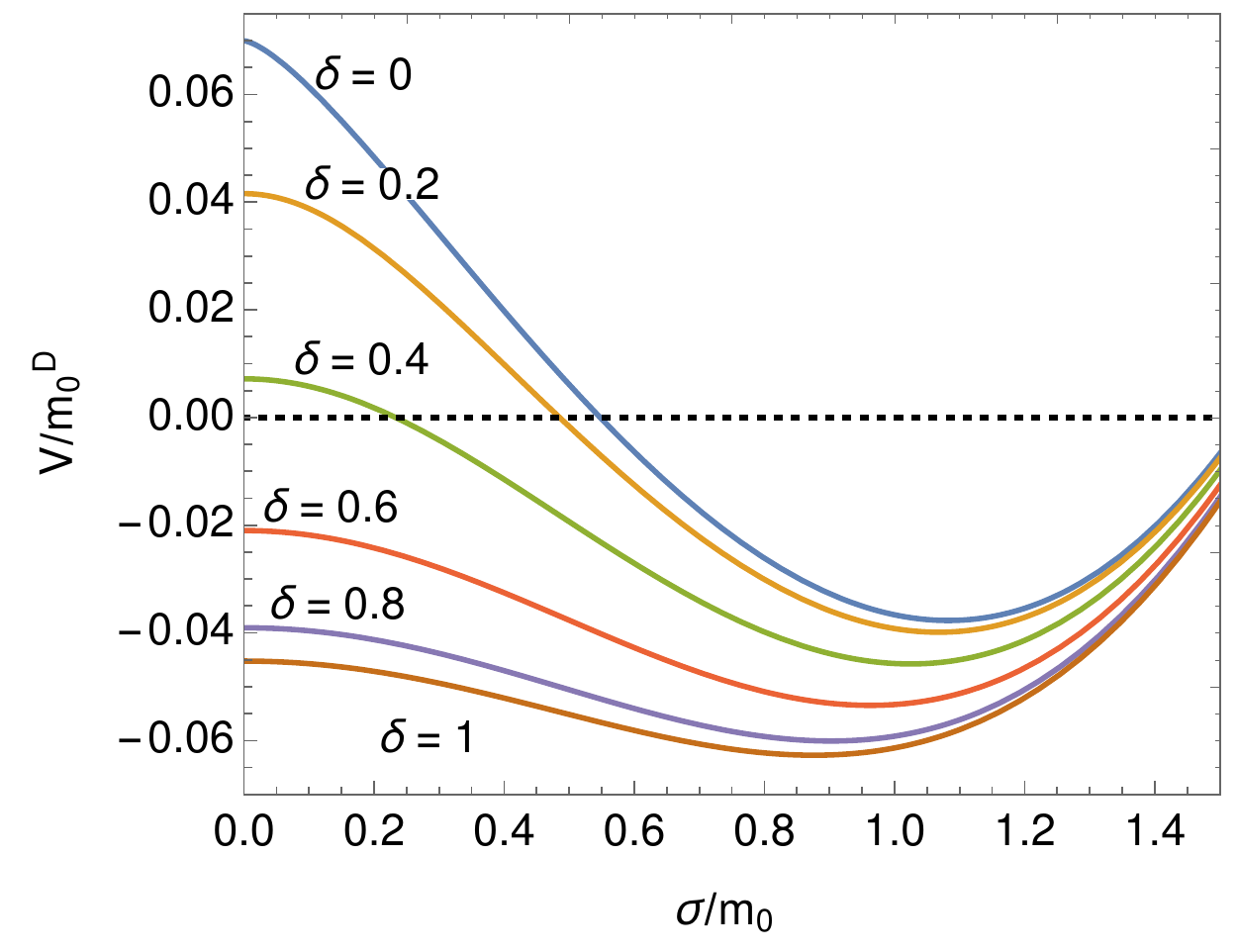}
     (b) $D=2.5$
    \end{center}
    \vglue 1mm
   \end{minipage}
   \\
   \begin{minipage}{0.35\hsize}
    \begin{center}
     \includegraphics[width=0.95\hsize,clip]
     {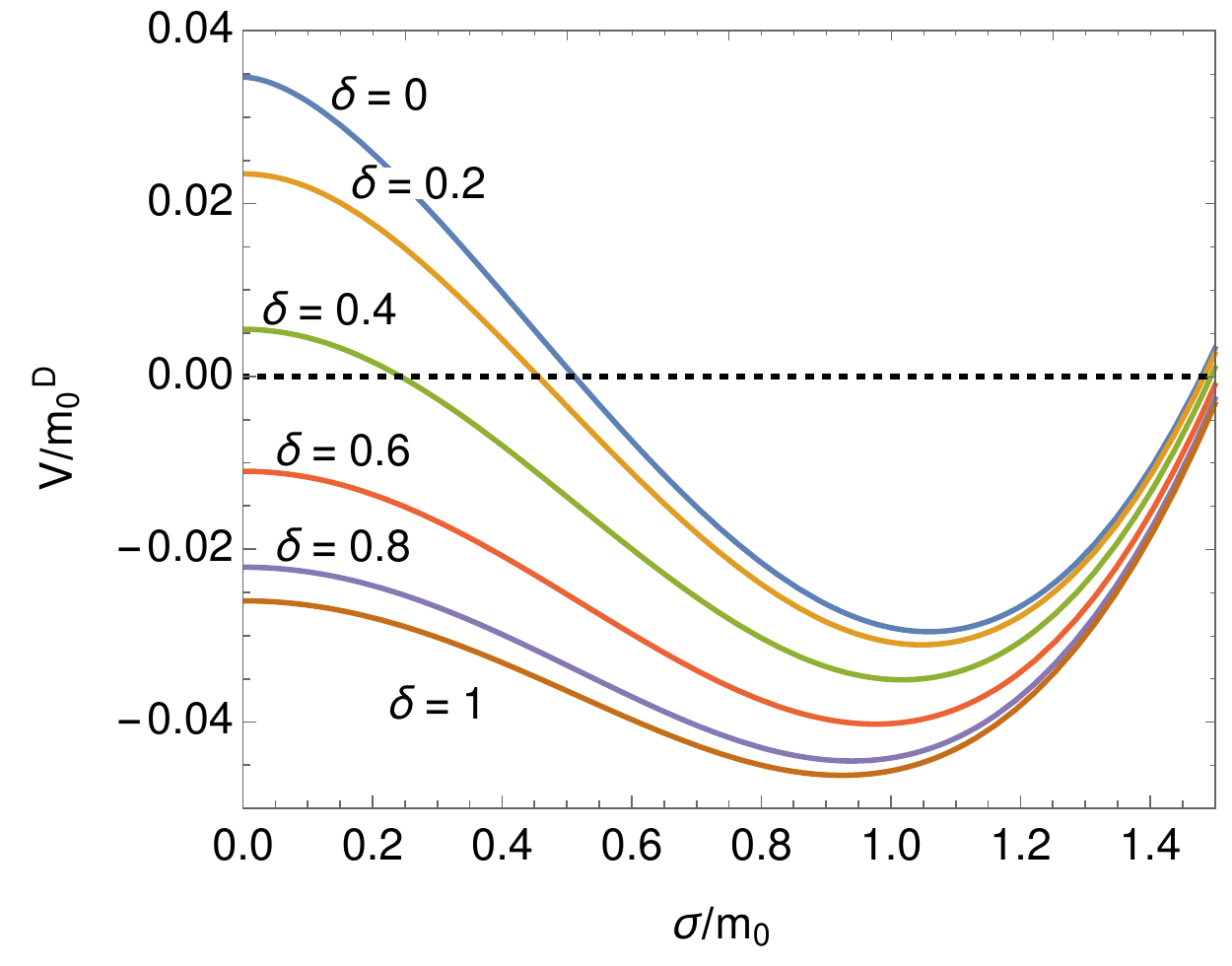}
     (c) $D=3$
    \end{center}
   \end{minipage}
&
    \begin{minipage}{0.35\hsize}
     \begin{center}
      \includegraphics[width=0.95\hsize,clip]
      {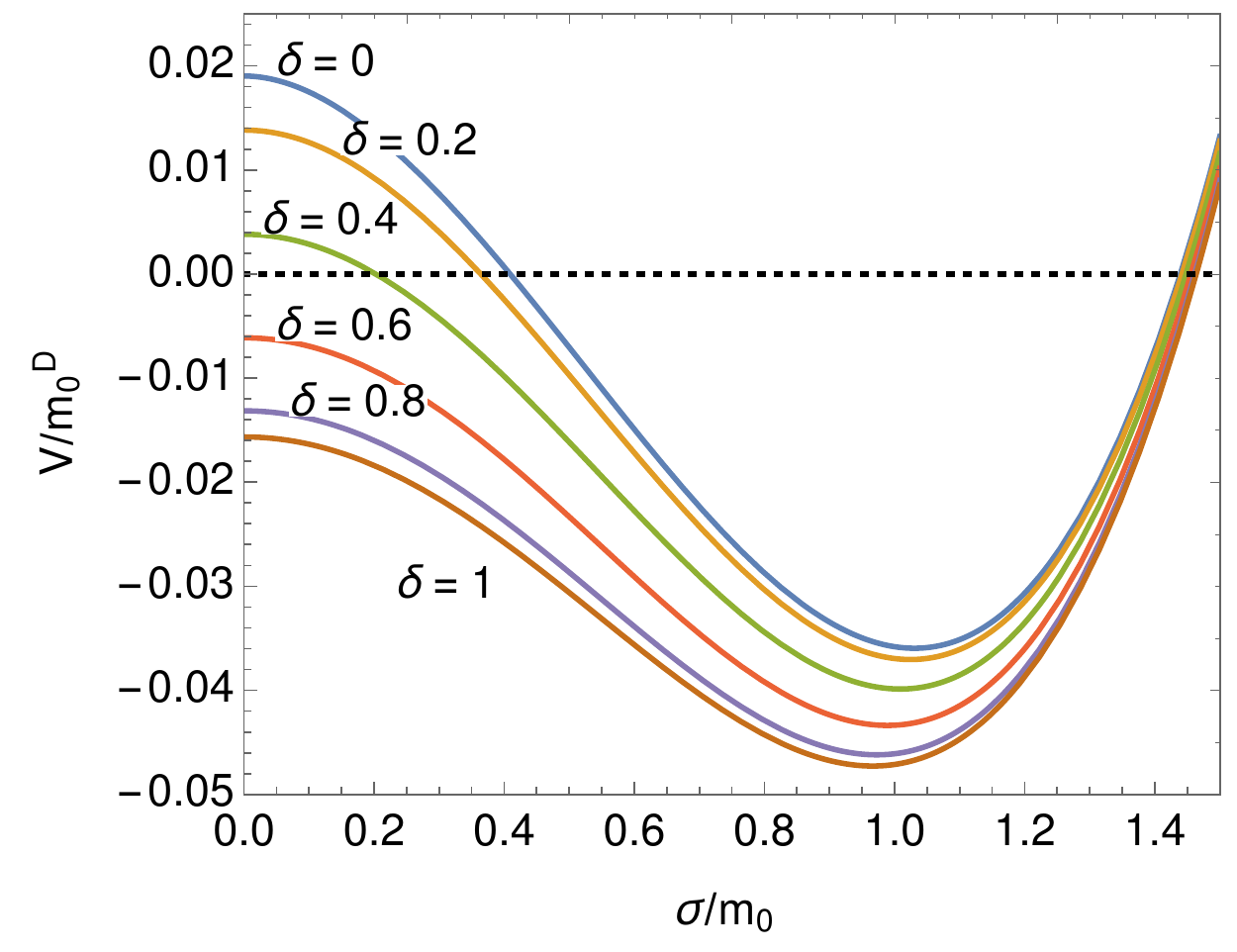}
      (d) $D=3.5$
     \end{center}
    \end{minipage}
  \end{tabular}
   \caption{Behavior of the effective potential on $\mathcal{M}^{D-1} \otimes S^1$ at $Lm_0=2.5$ for $\lambda_r >\lambda_{cr}$.}
   \label{EFFinite_Mass}
 \end{center}
\end{figure}

\begin{figure}[htbp]
 \begin{center}
  \begin{tabular}{cc}
   \begin{minipage}{0.35\hsize}
    \begin{center}
     \includegraphics[width=1\hsize]
     {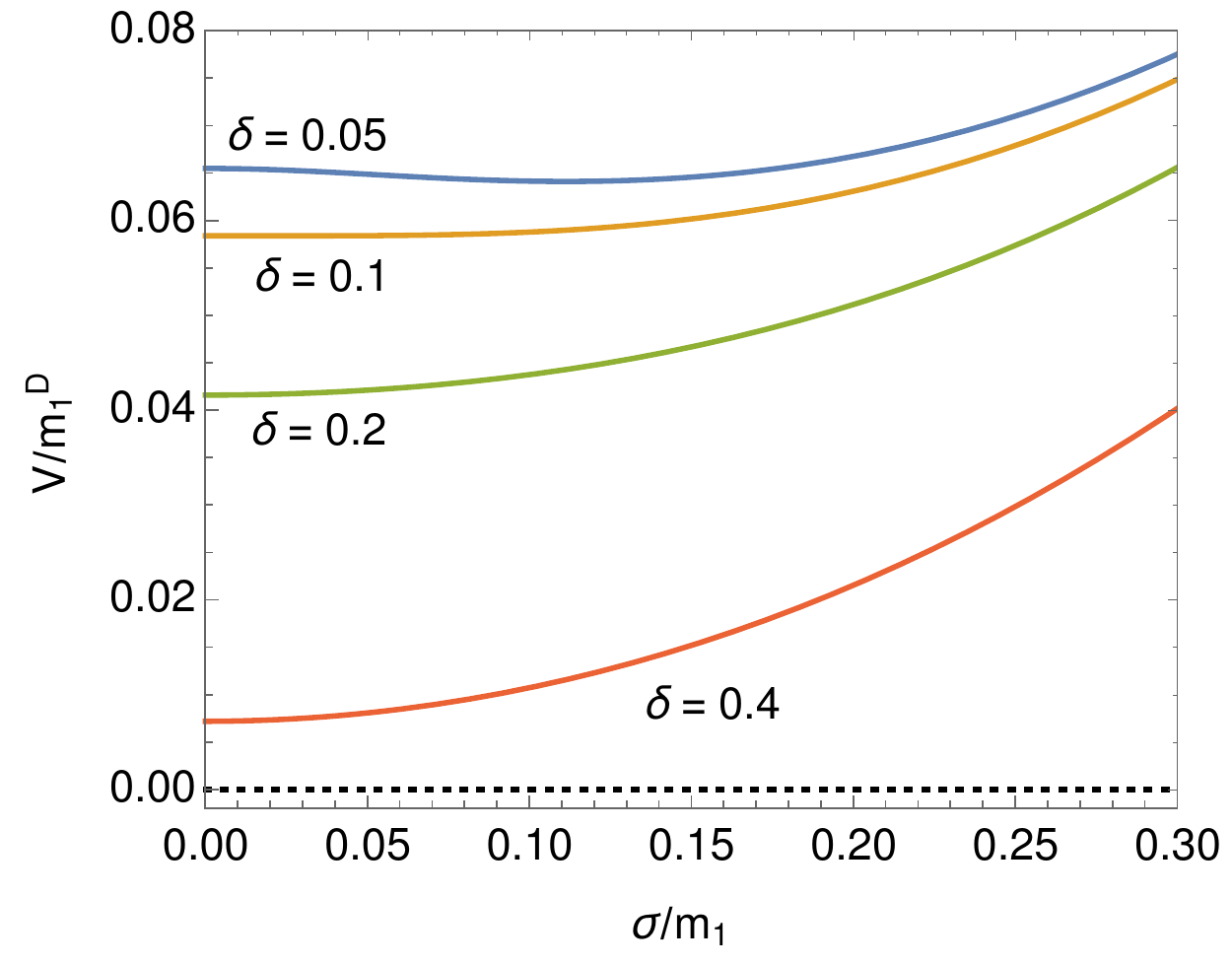}
      (a) $D=2.5$
    \end{center}
    \vglue 1mm
   \end{minipage}
&
   \begin{minipage}{0.35\hsize}
    \begin{center}
     \includegraphics[width=1\hsize]
     {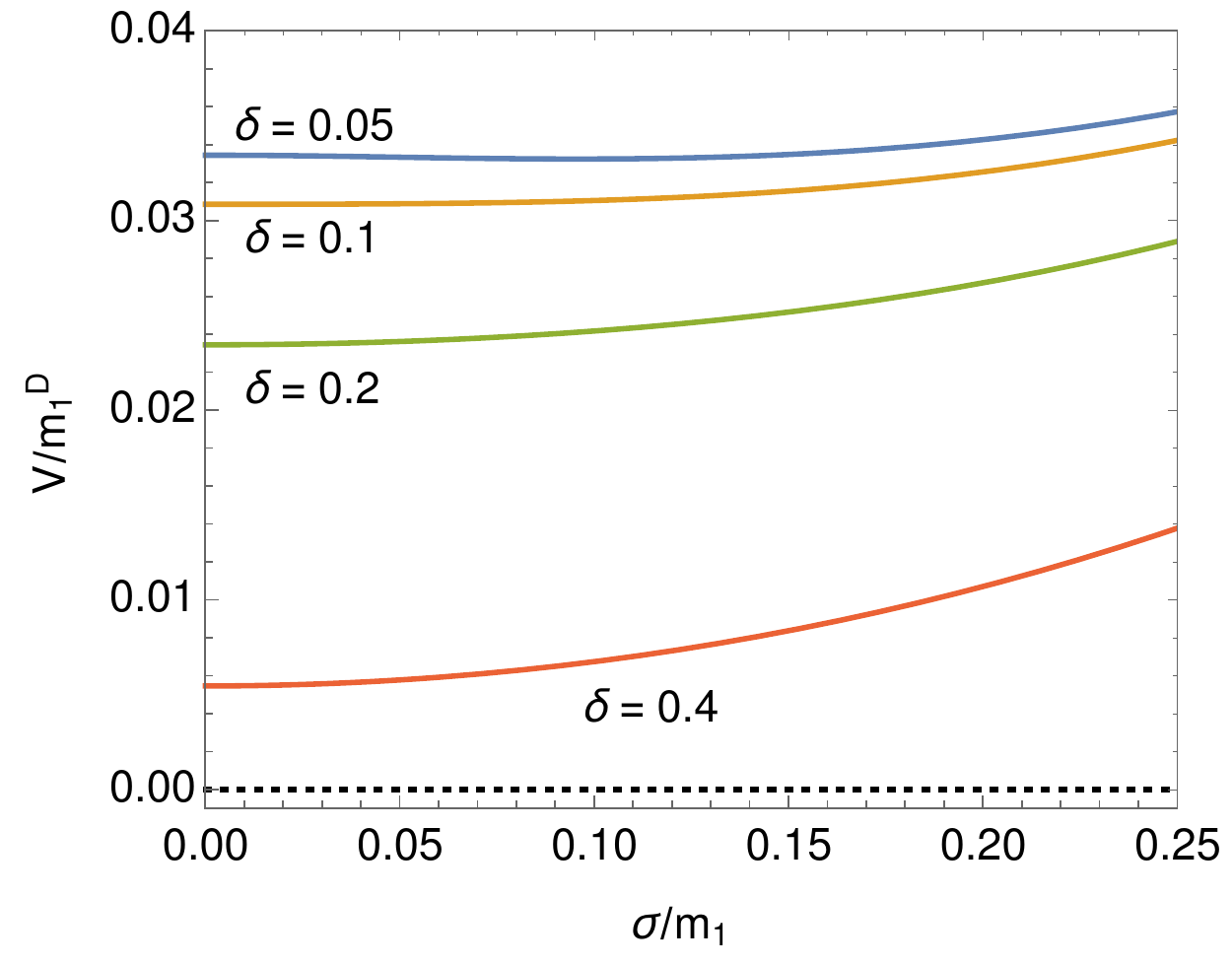}
     (b) $D=3$
    \end{center}
    \vglue 1mm
   \end{minipage}
\\
   \begin{minipage}{0.35\hsize}
    \begin{center}
     \includegraphics[width=1\hsize]
     {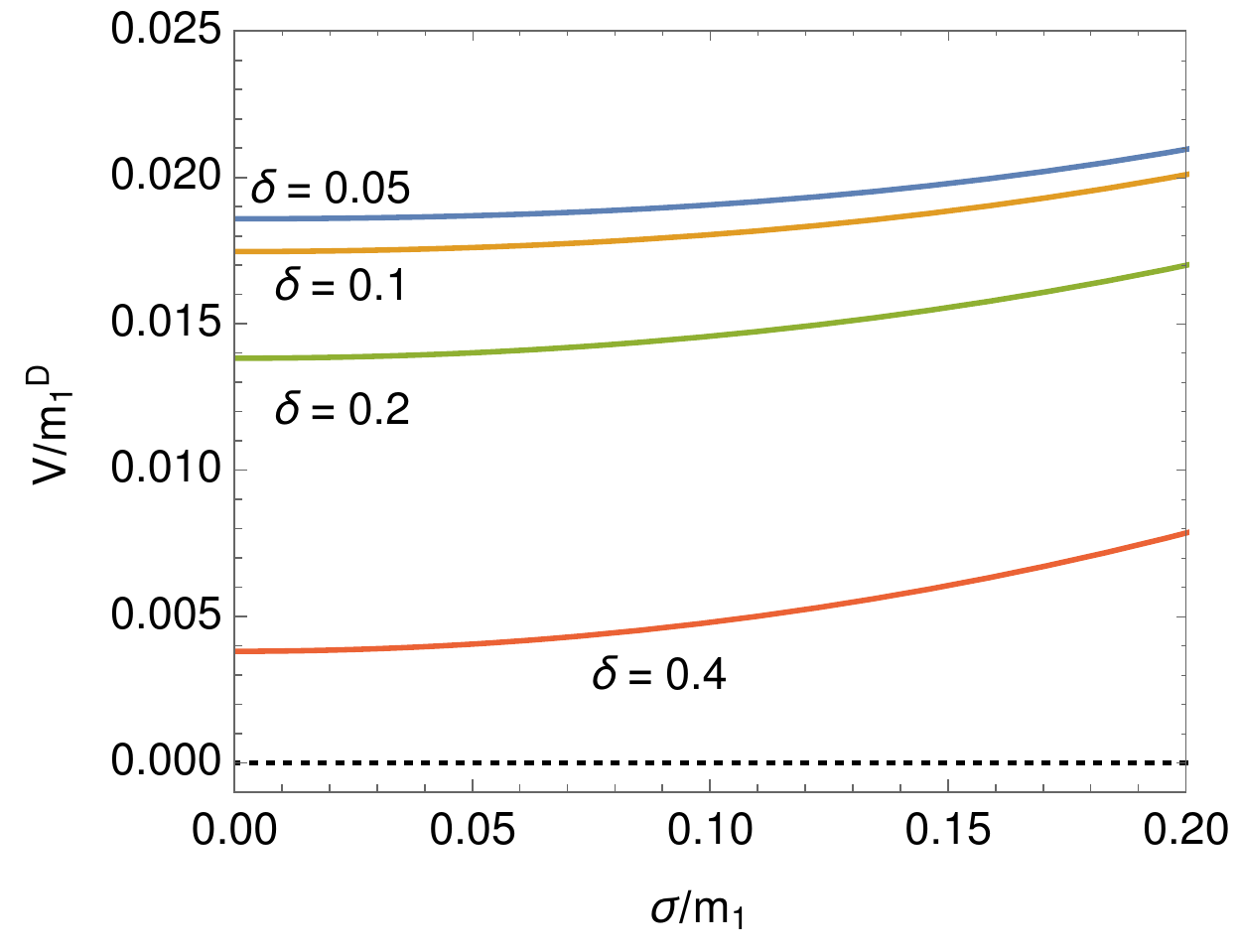}
     (c) $D=3.5$
    \end{center}
   \end{minipage}
&
   \begin{minipage}{0.35\hsize}
   \end{minipage}
  \end{tabular}
  \caption{Behavior of the effective potential on $\mathcal{M}^{D-1} \otimes S^1$ at $Lm_1=2.5$ for $\lambda_r <\lambda_{cr}$.}
   \label{EFFinite_MassLike}
 \end{center}
\end{figure}

Here we focus on the $\delta$ dependence and numerically evaluate the effective potential. It is known that the broken chiral symmetry is restored by the finite size effect for the anti-periodic boundary condition, $\delta=1$ and the finite size effect enhances the chiral symmetry breaking for the periodic boundary condition, $\delta=0$ \cite{Kim:1994es, Inagaki:1997kz, Ishikawa:1998uu}. Typical behavior of the effective potential is shown in Figs.~\ref{EFFinite_Mass} \ref{EFFinite_MassLike} for fixed lengths, $Lm_0 = 2.5$ and $Lm_1 = 2.5$. Because of the periodicity of the effective potential, $\tilde{V}(\sigma)|_{\delta} = \tilde{V}(\sigma)|_{\pm\delta+2 l \pi}$ for an arbitrary integer $l$, it is enough to study the effective potential within the interval, $0 \leq \delta \leq1$.

\begin{figure}[htbp]
    \begin{center}
     \begin{tabular}{cc}
      \begin{minipage}{0.4\hsize}
       \begin{center}
	\includegraphics[width=1\hsize]{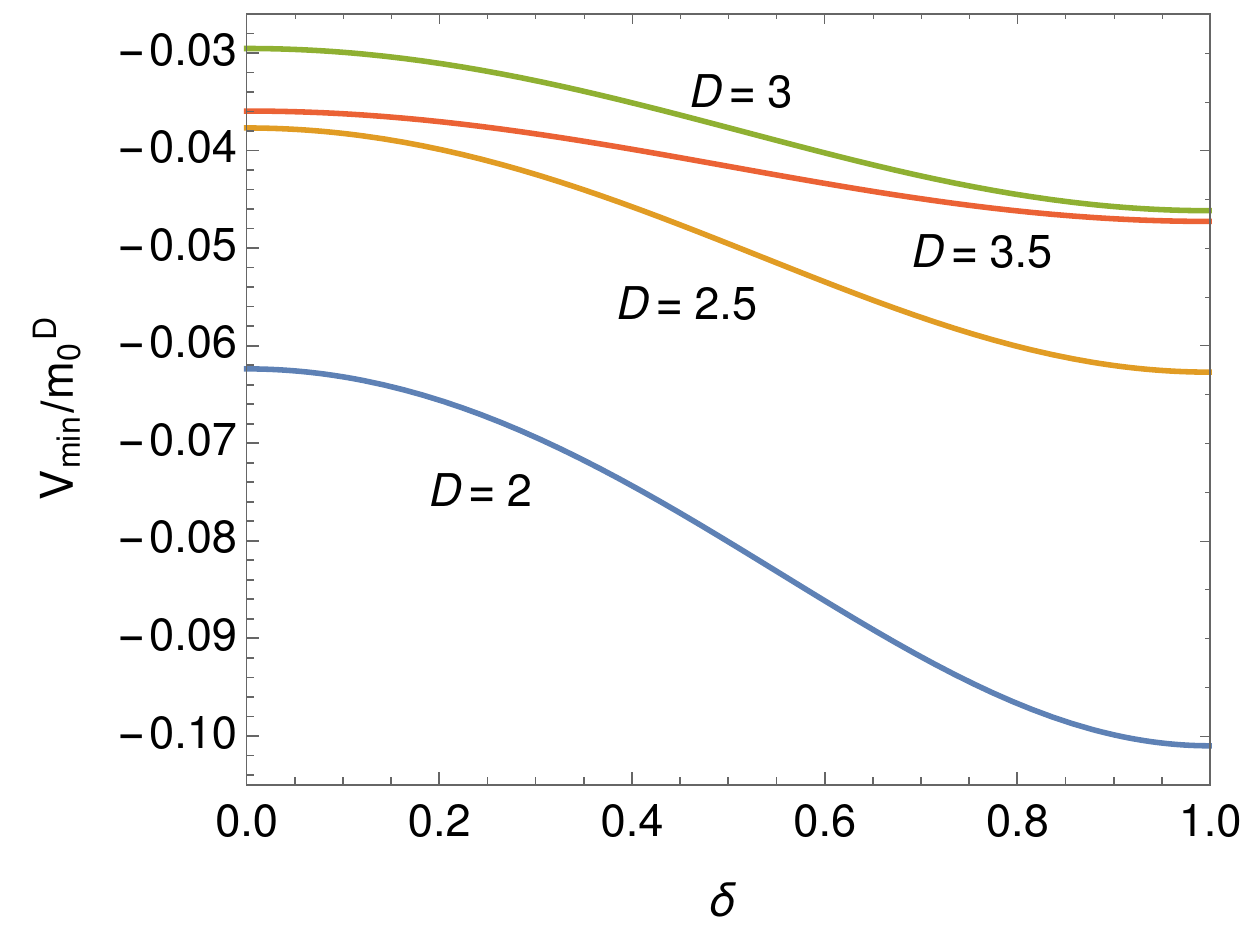}
       (a) $\lambda_r >\lambda_{cr}$, $Lm_0=2.5$
       \end{center}
      \end{minipage}
      &
      \begin{minipage}{0.4\hsize}
       \begin{center}
	\includegraphics[width=1\hsize]{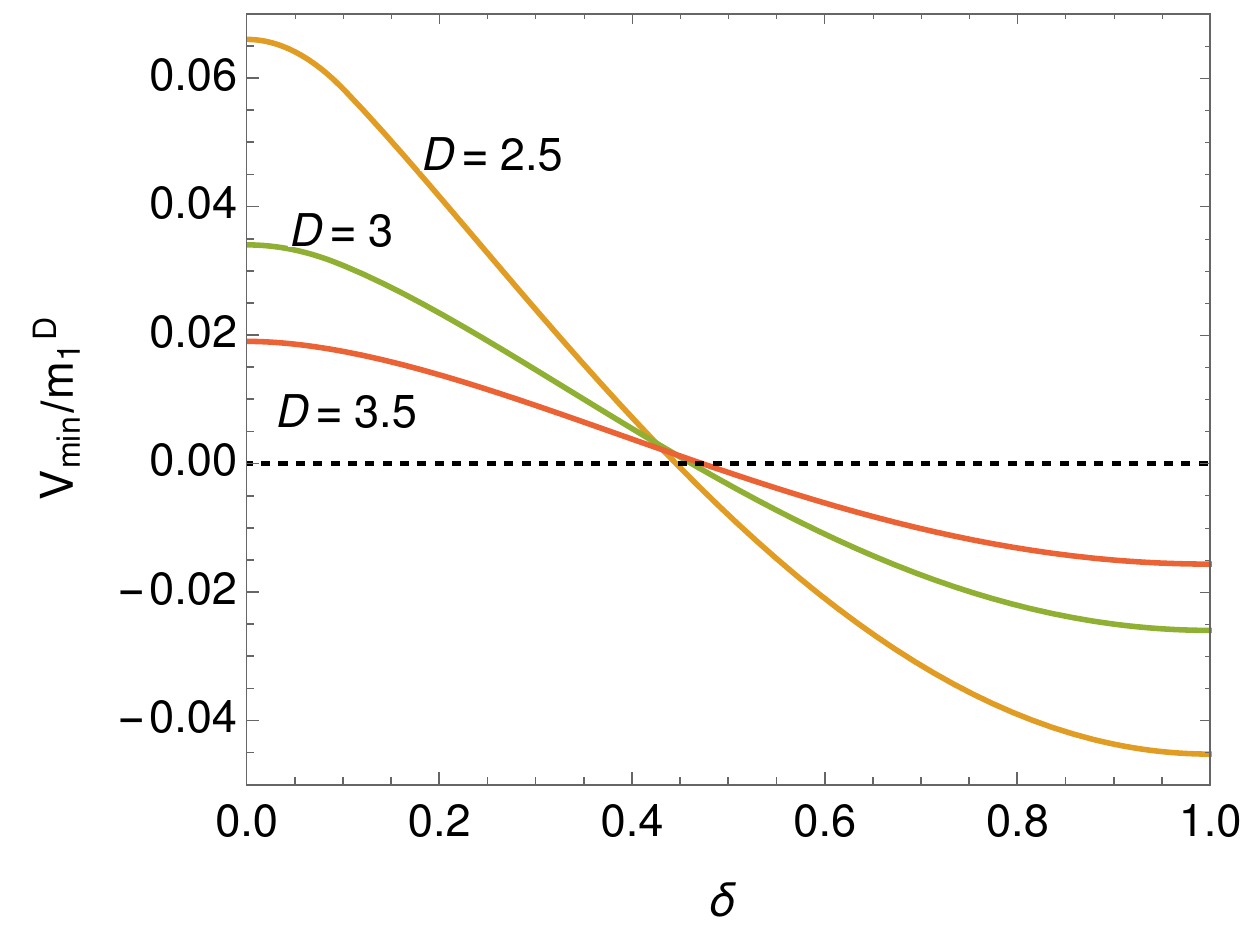}
       (b) $\lambda_r <\lambda_{cr}$, $Lm_1=2.5$
       \end{center}
      \end{minipage}
      \end{tabular}
    \end{center}
      \caption{Value of the effective potential at the minimum.}
     \label{Minimum_ML}
   \end{figure}

In Fig.~\ref{EFFinite_Mass} it is observed that the effective potential has a non-trivial minimum for any dimensions. The non-trivial minimum shows the existence of the ground state which breaks the chiral symmetry. In Fig.~\ref{EFFinite_MassLike} the broken phase is observed at $(D,\delta) = (2.5, 0.05), (3, 0.05)$ around the periodic boundary condition. As is shown in Figs.~\ref{EFFinite_Mass}, \ref{EFFinite_MassLike}, the value of the effective potential at the minimum, $V_{min}$, decreases with the phase, $\delta$, approaching the anti-periodic boundary condition, $\delta=1$.  We plot the minimum value, $V_{min}$, as a function of the $\mathrm{U}(1)$ phase $\delta$ in Fig.~\ref{Minimum_ML}. If the $\mathrm{U}(1)$ phase $\delta$ is a dynamical valuable, the stable state is found at the anti-periodic boundary condition, $\delta=1$.  

\begin{figure}[tbp]
 \begin{center}
  \begin{tabular}{cc}
   \begin{minipage}{0.4\hsize}
       \begin{center}
	\includegraphics[width=1\hsize]{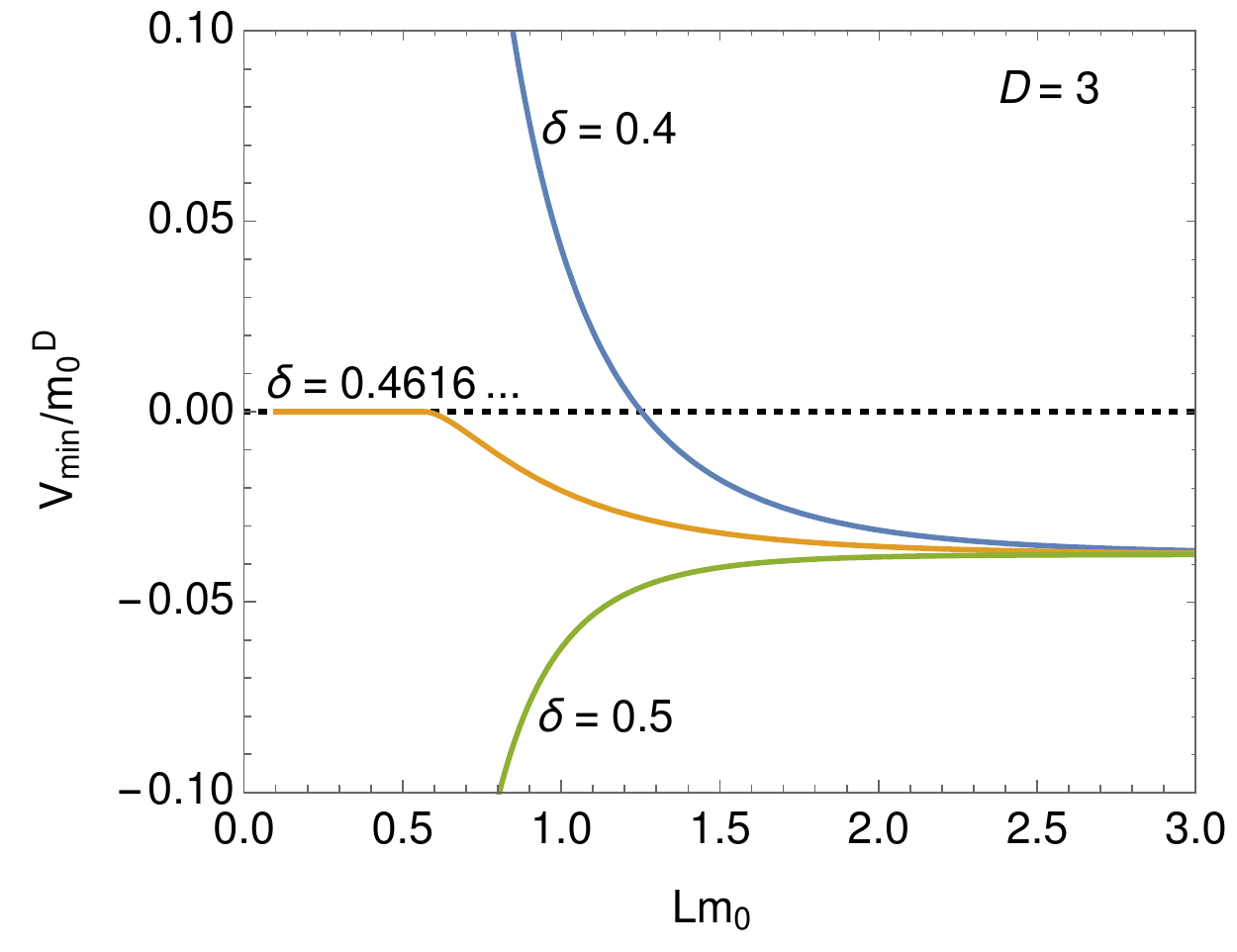}
	{(a) $\lambda_r>\lambda_{cr}$}
       \end{center}
   \end{minipage}
&
   \begin{minipage}{0.4\hsize}
    \begin{center}
     \includegraphics[width=1\hsize]{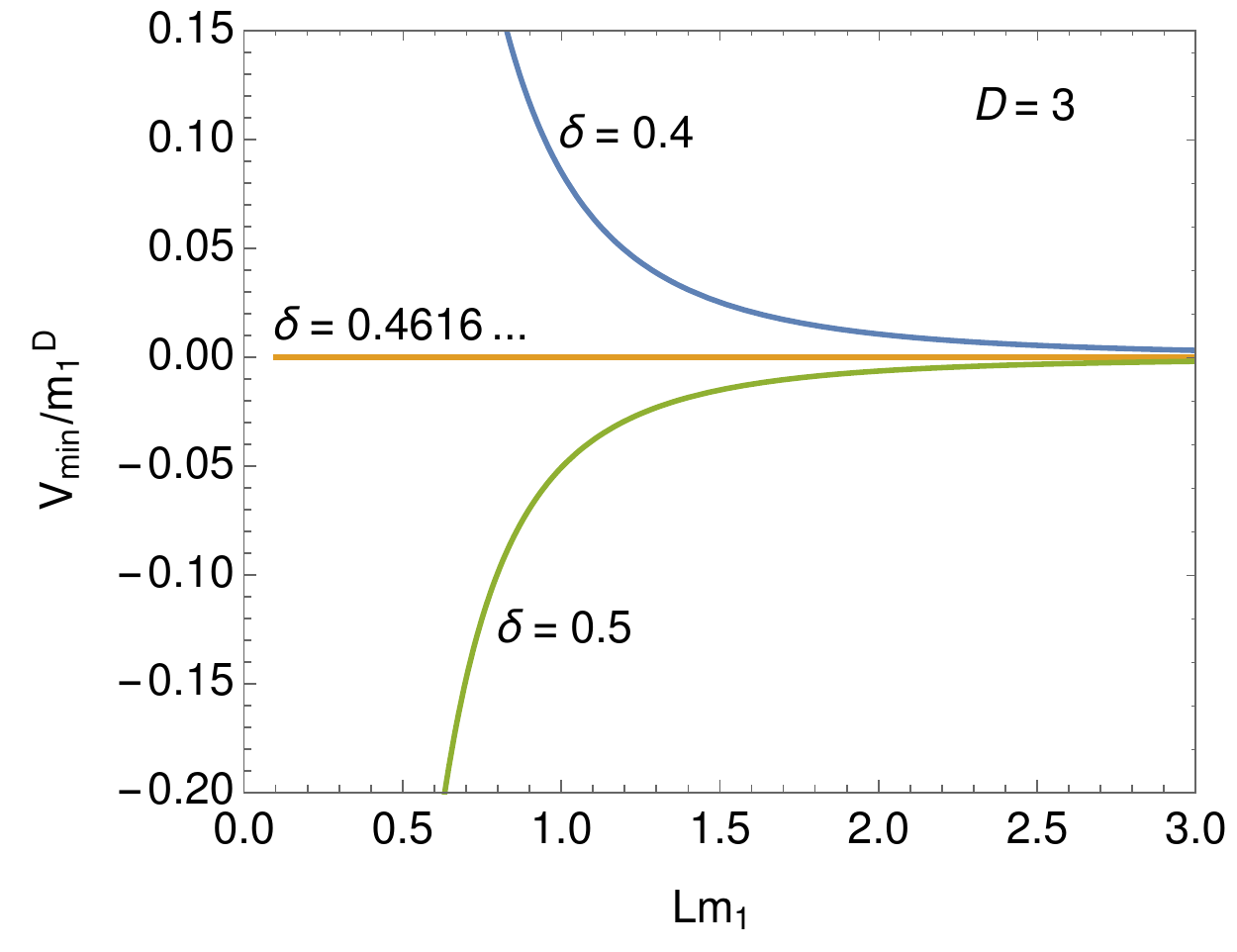}
     {(b) $\lambda_r <\lambda_{cr}$}
    \end{center}
   \end{minipage}
  \end{tabular}
  \caption{Value of the effective potential at the minimum for $D=3$.}
  \label{Minimum_Length}
 \end{center}
\end{figure}

In Fig.~\ref{Minimum_Length} the minimum value, $V_{min}$, is plotted as a function of the length of the compactified space, $L$. In the figure we show the typical behavior of the minimum for $D=3$. It monotonically increases and decreases for a small and large $\delta$, respectively. For a specific phase ($\delta=0.46\dots$ for $D=3$) the minimum value vanishes for $L<L_{cr}$ in which the chiral symmetry is restored. The $L$-dependence of $V_{min}$ induces the Casimir force as is discussed in Sec.~4.

The dynamically generated fermion mass, $m$, is given by the field value, $\sigma$, at the minimum of the effective potential. It is obtained as a non-vanishing solution of the gap equation,
\begin{align}
 \left.\frac{\partial \tilde{V}(\sigma)}{\partial \sigma}\right|_{\sigma=m} = 0. 
 \label{gap}
\end{align}
Substituting equation \eqref{EFSumInt} into \eqref{gap}, we derive
\begin{align}
 & \frac{1}{2\sqrt{\pi}} \Gamma\left( 1 - \frac{D}{2}\right)\Gamma\left(\frac{D-1}{2}\right)
 \left[(-1)^\alpha - \left( \frac{m^2}{m_\alpha^2}\right)^{\frac{D}{2} -1}\right]   \notag\\
 & \hspace{4ex} =  -
 \int_0^\infty \frac{\dmeasure{}{K}}{\sqrt{K^2 + m^2}} \left(\frac{K}{m_\alpha}\right)^{D-2}
 \frac{\exp\left(-L \sqrt{K^2 +m^2} \right)- \cos \left( \pi\delta \right)}{\cosh \left(L \sqrt{K^2 +m^2}\right) -\cos\left(\pi\delta\right)} .
 \label{gap:int}
\end{align}
\begin{figure}[tpb]
 \begin{center}
  \begin{tabular}{cc}
      \begin{minipage}{0.4\hsize}
       \begin{center}
	\includegraphics[width=1\hsize]{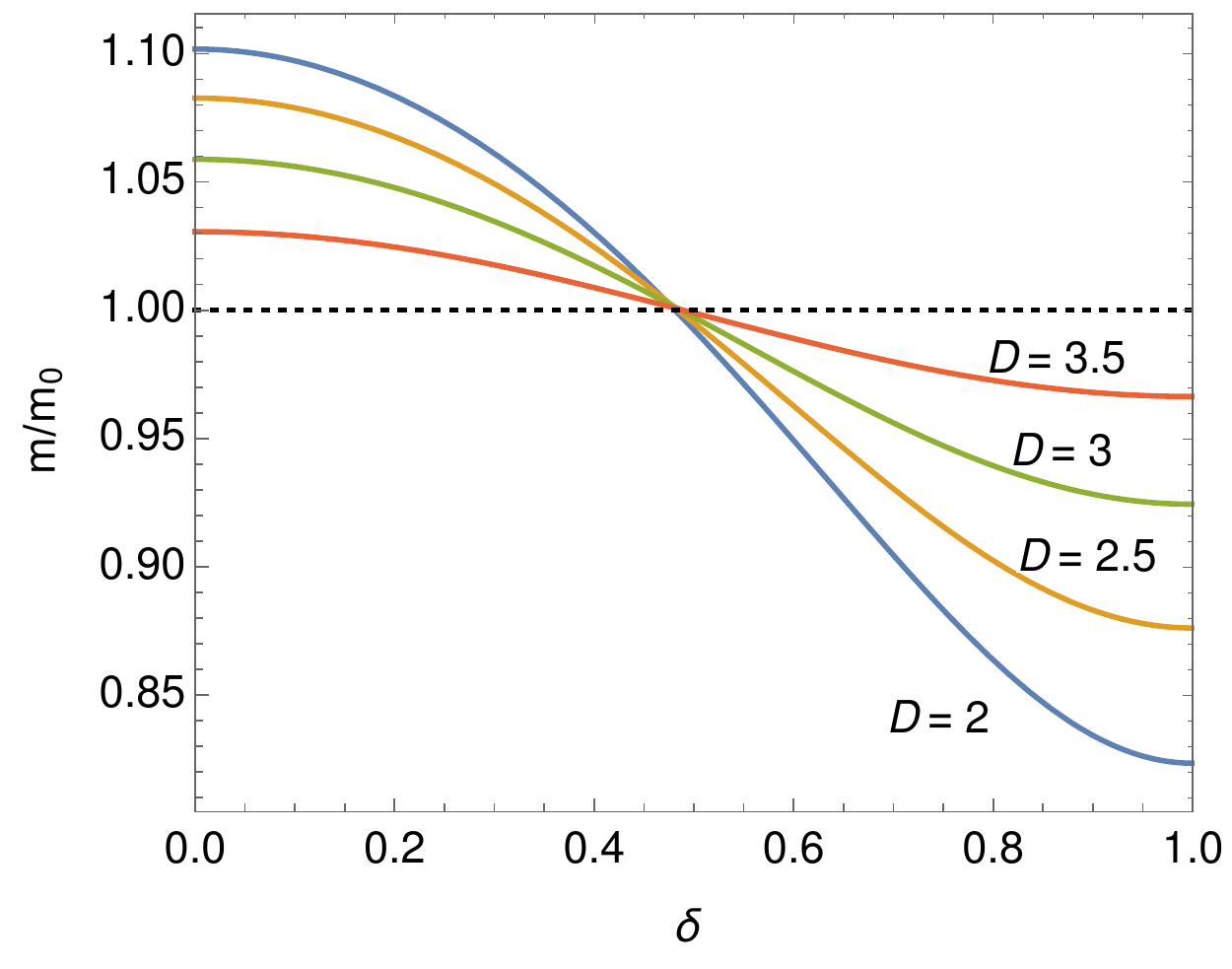}
	(a) $\lambda_r>\lambda_{cr}$, $Lm_0=2.5$
       \end{center}
      \end{minipage}
   &
   \begin{minipage}{0.4\hsize}
    \begin{center}
     \includegraphics[width=1\hsize]{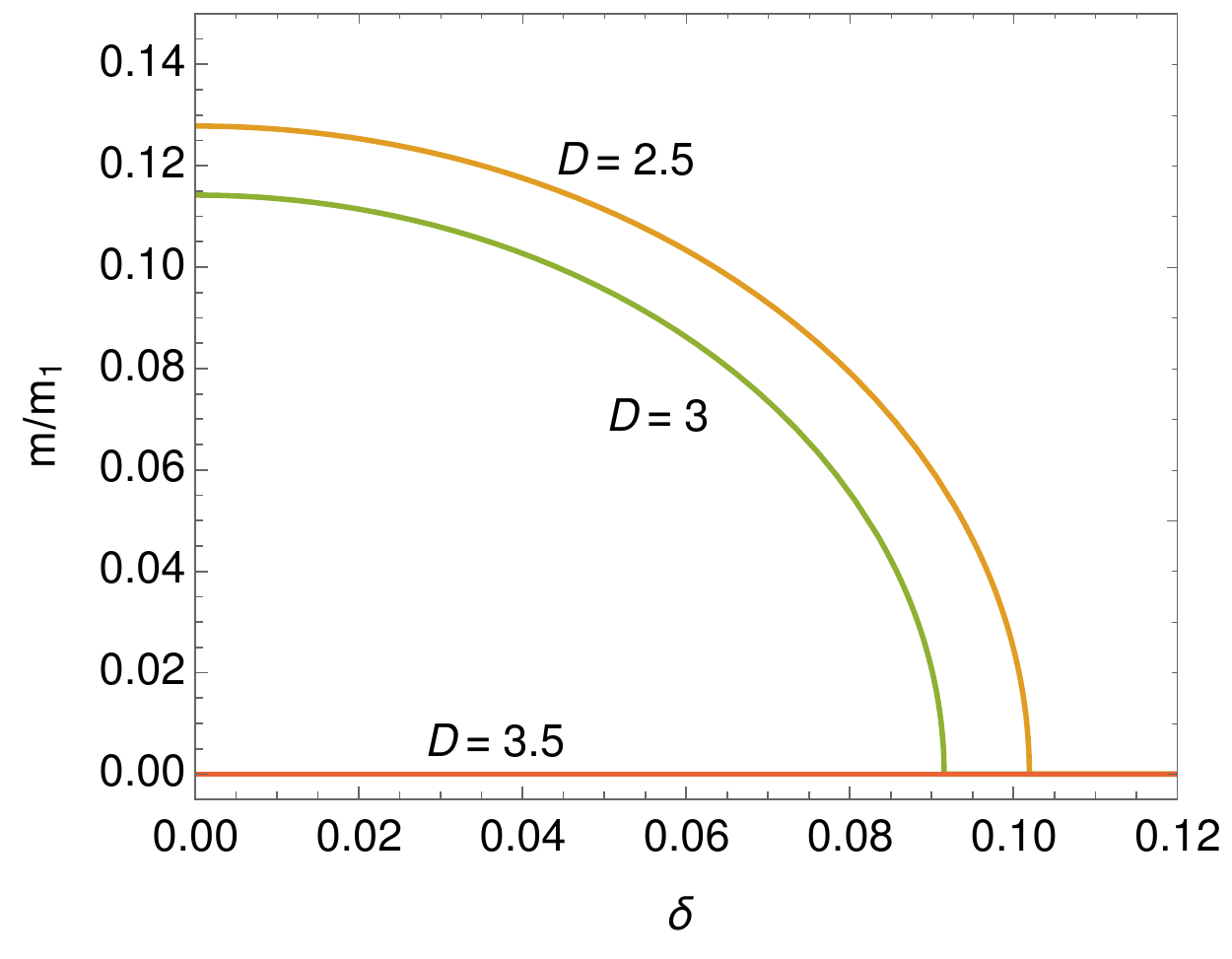}
     (b) $\lambda_r <\lambda_{cr}$, $Lm_1=2.5$
    \end{center}
   \end{minipage}
  \end{tabular}
 \end{center}
 \caption{Dynamically generated fermion mass.}
 \label{Minimum_DGMass_ML}
\end{figure}

\begin{figure}[tbp]
 \begin{center}
  \begin{tabular}{cc}
   \begin{minipage}{0.4\hsize}
    \begin{center}
     \includegraphics[width=1\hsize]{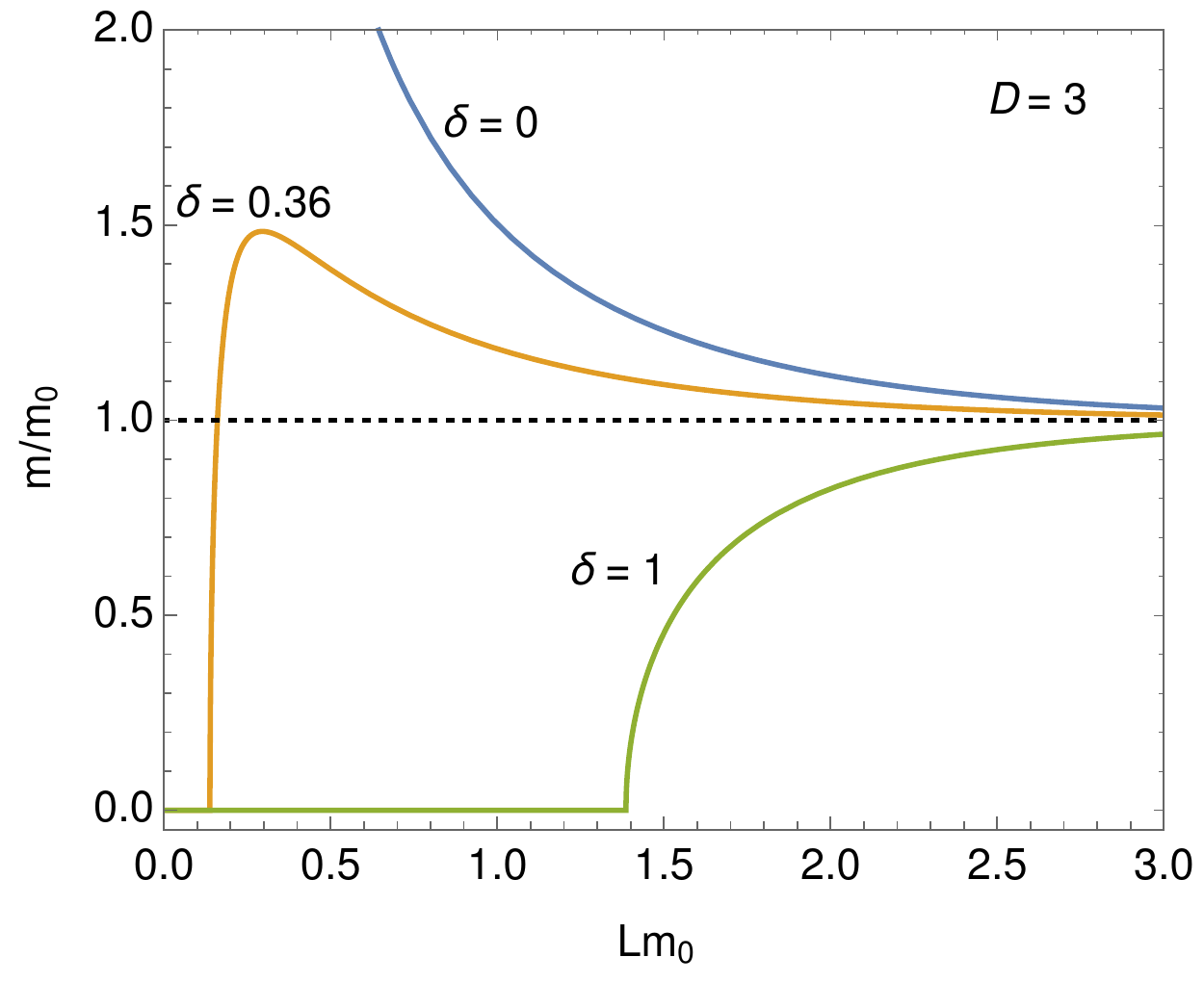}
     {(a) $\lambda_r>\lambda_{cr}$}
    \end{center}
   \end{minipage}
   &
   \begin{minipage}{0.4\hsize}
    \begin{center}
     \includegraphics[width=1\hsize]{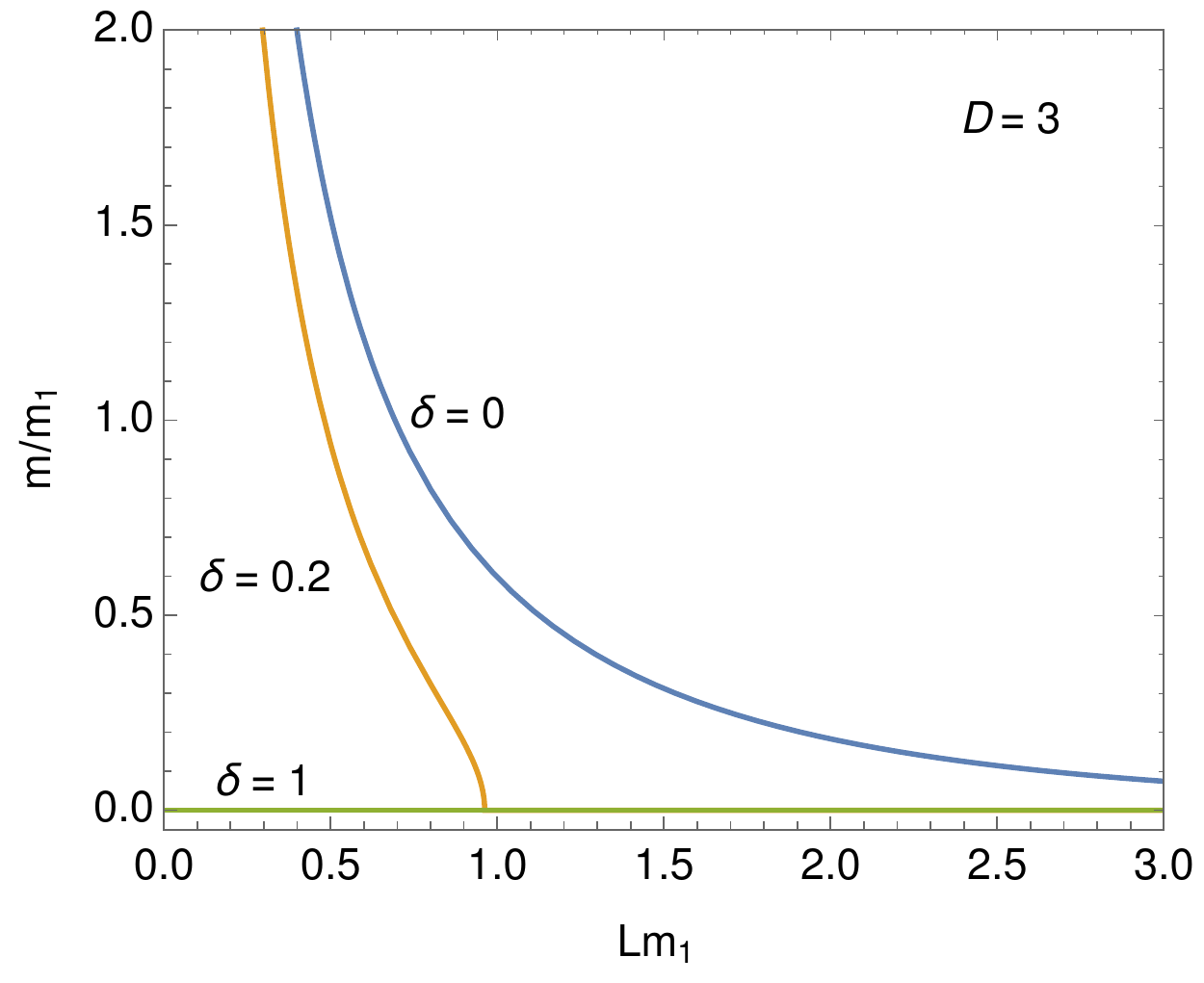}
     {(b) $\lambda_r <\lambda_{cr}$}
    \end{center}
   \end{minipage}
  \end{tabular}
  \caption{Dynamically generated fermion mass for $D=3$.}
  \label{DGMass_Length}
 \end{center}
\end{figure}
We numerically solve equation \eqref{gap:int} and draw the behavior of the dynamically generated mass as a function of $\delta$ for fixed lengths in Fig.~\ref{Minimum_DGMass_ML}. It is clearly seen that the generated mass, $m$, monotonically decreases as the phase, $\delta$, approaches the unity, $\delta\to 1$. In the weak coupling case, $\lambda_r < \lambda_{cr}$, with $Lm_1=2.5$ it is observed that the fermion mass disappears above a critical value of $\delta$ and  the second order phase transition takes place for $D=2.5$ and $3$. Only the symmetric phase, $m=0$, is observed for $D=3.5$. 
In Fig.~\ref{DGMass_Length} the generated mass is plotted as a function of the length, $L$, for typical $\delta$ at $D=3$. The finite size effect restores the broken chiral symmetry  for $\delta=1$ and enhances the chiral symmetry breaking for $\delta=0$. For intermediate $\delta$ combined behavior is observed. We checked that the finite size phase transition is always of the second order.

\begin{figure}[tbp]
 \begin{center}
  \begin{tabular}{cc}
   \begin{minipage}{0.4\hsize}
    \begin{center}
     \includegraphics[width=1\hsize]{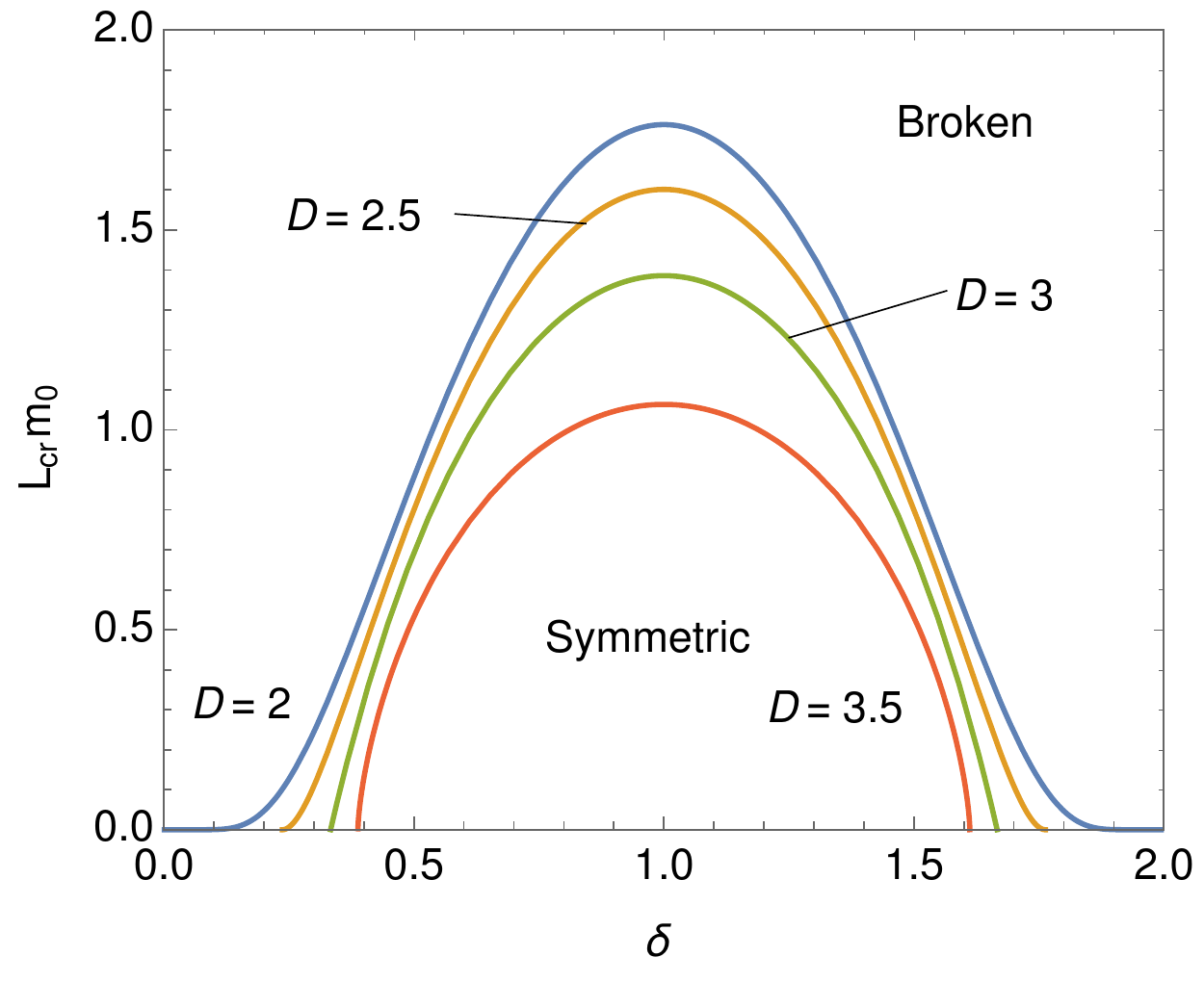}
     {(a) $\lambda_r>\lambda_{cr}$}
    \end{center}
   \end{minipage}
&
   \begin{minipage}{0.4\hsize}
    \begin{center}
     \includegraphics[width=1\hsize]{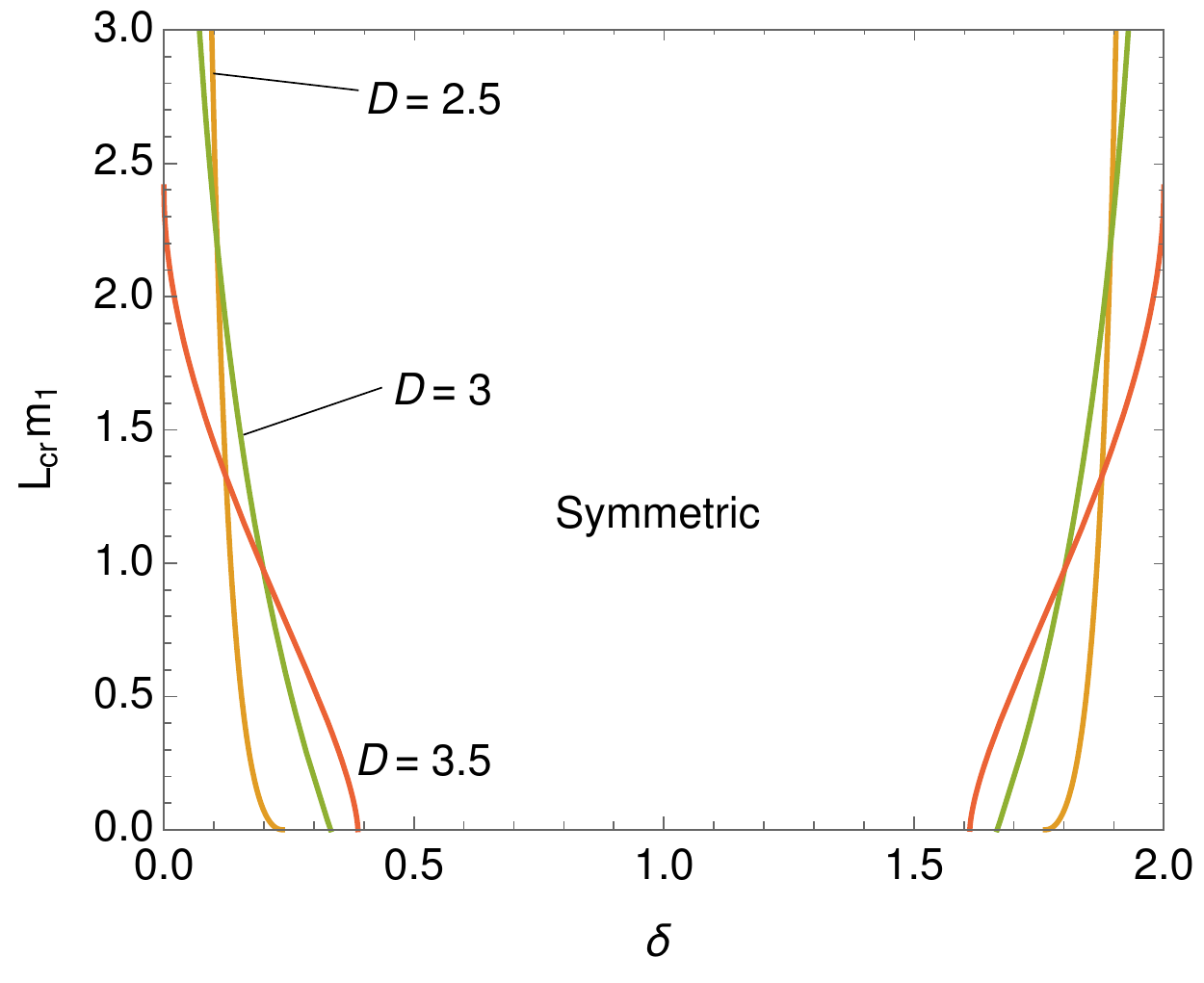}
     {(b) $\lambda_r <\lambda_{cr}$}
    \end{center}
   \end{minipage}
  \end{tabular}
  \caption{Phase structure on $\delta-L$ plane. The chiral symmetry is broken above the lines for $\lambda_r >\lambda_{cr}$ and below  the lines for $\lambda_r<\lambda_{cr}$.}
  \label{CriticalLength}
 \end{center}
\end{figure}

Since the phase transition is of the second order, the phase boundary is found at the massless limit, $m\to 0$, of the gap equation. We adopt the expression \eqref{Appendix:epot2} and derive the explicit expression for the critical length, $L_{cr}$, as a function of the phase, $\delta$. Differentiating equation \eqref{Appendix:epot2} with respect to $\sigma$, the divergent function, $C(L)$, is dropped and the gap equation reads 
\begin{align}
 L m_\alpha = 2 \sqrt{\pi} \frac{(-1)^\alpha }{m_\alpha^{D-3}}\frac{\Gamma\left(\frac{3-D}{2}\right)}{\Gamma\left(1-\frac{D}{2}\right)} 
 \sum_{n=-\infty}^\infty \left( \omega_n^2 + m^2\right)^{\frac{D-3}{2}} . 
 \label{gap:sum}
\end{align}
The critical length, $L_{cr}$, which divides the symmetric and broken phases is obtained by taking the massless limit, $m\to 0$, of \eqref{gap:sum}, 
At the limit the summation in \eqref{gap:sum} is described by the Hurwitz zeta function, $\zeta(z,a)$, which is defined in \eqref{def:zeta} and the critical length, $L_{cr}$, is given by
\begin{align}
 L_{cr}m_\alpha =
 \left\{
 \begin{aligned}
  &2\pi \left(\frac{(-1)^\alpha}{\sqrt{\pi}}\frac{\Gamma\left(\frac{3-D}{2}\right)}{\Gamma\left(1-\frac{D}{2}\right)}  \left[ \zeta\left(3-D,1-\frac{\delta}{2}\right) + \zeta\left( 3-D,\frac{\delta}{2}\right)\right]\right)^{\frac{1}{D-2}}, &0<\delta <2,
  \\
  &2\pi \left(\frac{2(-1)^\alpha}{\sqrt{\pi}}\frac{\Gamma\left(\frac{3-D}{2}\right)}{\Gamma\left(1-\frac{D}{2}\right)} \zeta\left(3-D\right) \right)^{\frac{1}{D-2}}, &\delta =0. 
 \end{aligned}
 \right.
 \label{eq:CriticalLength}
\end{align}
In the case of the anti-periodic condition, $\delta=1$, this formula coincides with the one for the critical temperature on ${\mathcal M}^D$ which is derived in \cite{Inagaki:1994ec}.

We numerically calculate the critical length and draw the phase diagram in Fig. \ref{CriticalLength}. For $\lambda_r >\lambda_{cr}$ the symmetric phase is observed around $\delta=1$ where the broken chiral symmetry is restored by the finite size effect. Only the broken phase is observed for the periodic boundary condition, $\delta=0$. In the weak coupling case  $\lambda_r <\lambda_{cr}$ the chiral symmetry is broken around $\delta=0$. 
The critical length is divergent at $\delta=0$ for $D=3$. As long as the length is finite $L<\infty$, the chiral symmetry is always broken in the periodic boundary condition, $\delta=0$, for $2< D\leq 3$. Since the finite effect vanishes at the limit $L\to\infty$, the symmetric phase appears at the $L\to\infty$ limit for $\delta=0$. The situation is also observed in the behavior of the fermion mass for the lines at $\delta=0$ in Fig.~\ref{DGMass_Length}. 

\begin{figure}[tbp]
 \begin{center}
  \begin{tabular}{cc}  
   \begin{minipage}{0.4\hsize}
    \begin{center}
     \includegraphics[width=1\hsize]{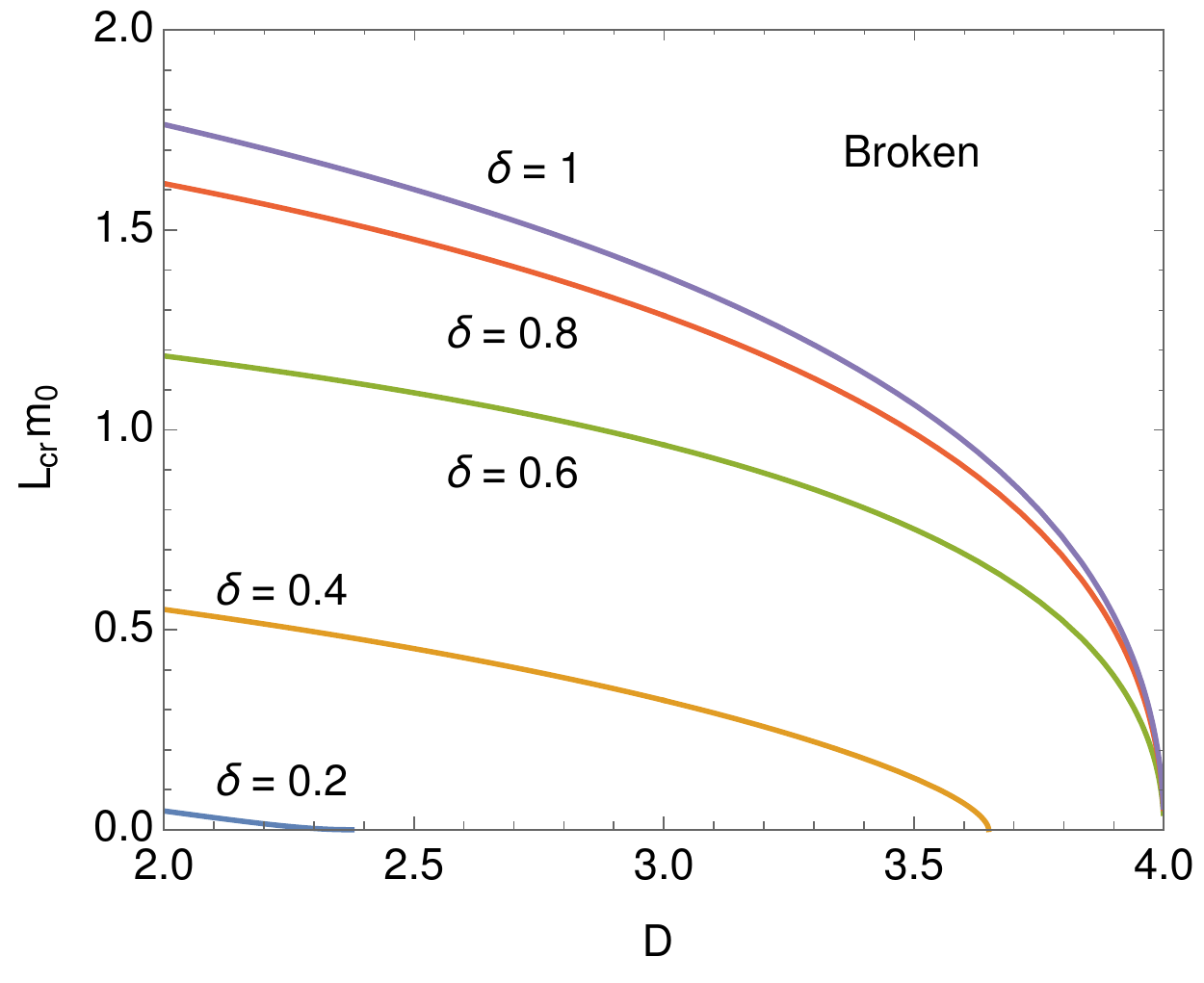}
     {(a) $\lambda_r>\lambda_{cr}$}
    \end{center}
   \end{minipage}
   &
   \begin{minipage}{0.4\hsize}
    \begin{center}
     \includegraphics[width=1\hsize]{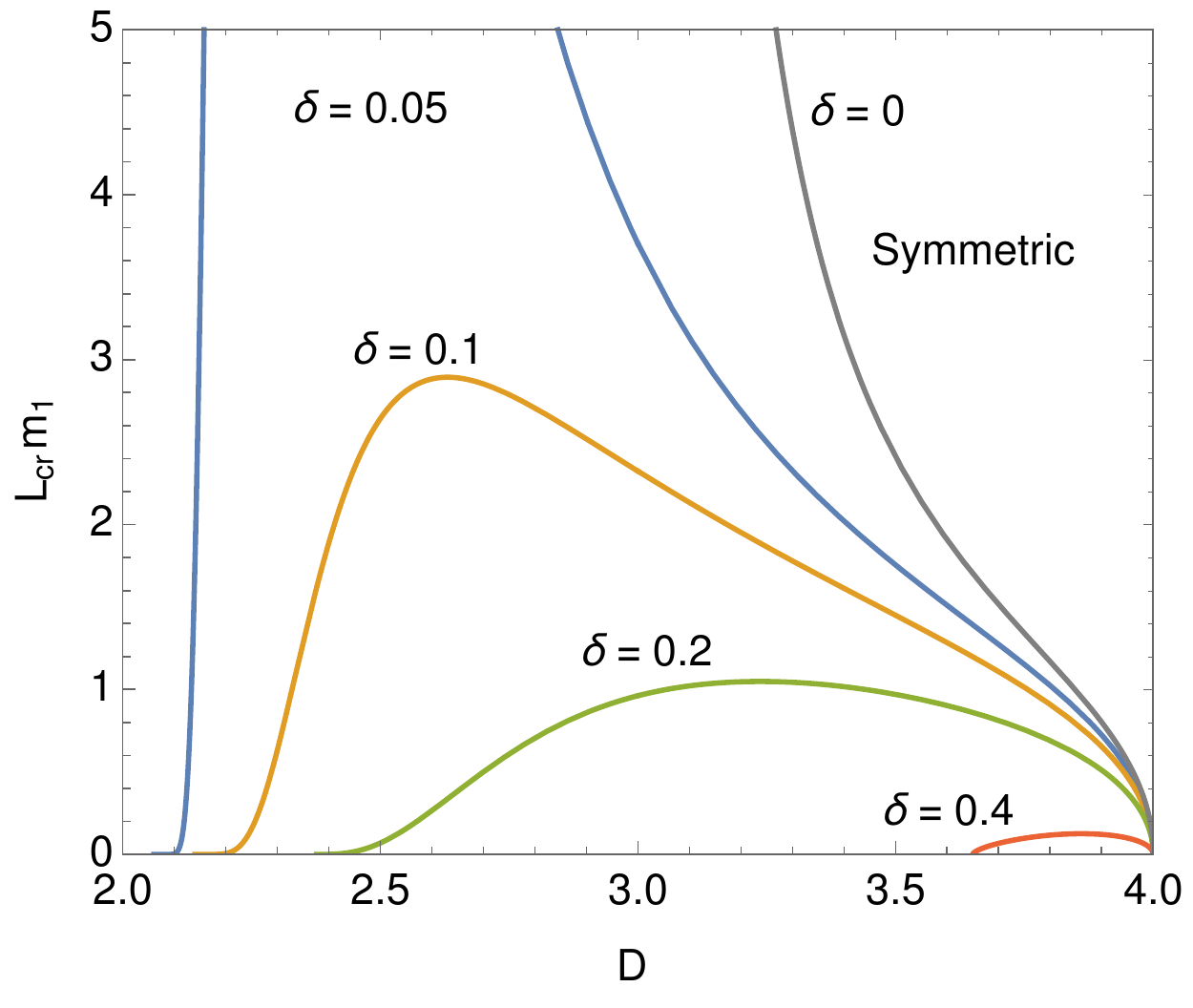}
     {(b) $\lambda_r<\lambda_{cr}$}
    \end{center}
   \end{minipage}
  \end{tabular}
  \caption{Phase structure on $D-L$ plane. The chiral symmetry is broken above the lines for $\lambda_r >\lambda_{cr}$ and below  the lines for $\lambda_r<\lambda_{cr}$.}
  \label{CriticalLength2}
 \end{center}
\end{figure}

The $D$-dependences of the chiral symmetry breaking are useful to understand the phase structure. As is shown in Fig.~\ref{CriticalLength2} (a), the chiral symmetry breaking is enhanced as the phase approaches from $\delta=1$ to $0$. It is clearly seen in Fig.~\ref{CriticalLength2} (b) that the broken phase appears around the periodic boundary condition, $\delta=0$, even in the weak coupling case $\lambda_r<\lambda_{cr}$. It is denoted that equation \eqref{eq:CriticalLength} at $\delta=0$ is defined only for $3\leq D <4$. In the case of the periodic boundary condition in lower dimensions $2\leq D<3$ the symmetric phase only realizes at the $L\to\infty$ limit.

 \section{Casimir effect}

In the previous section the effective potential on the ground state is described as a function of the size of the compactified spatial direction $L$ and the $\mathrm{U}(1)$ phase $\delta$ at the boundary. As is known as the Casimir effect \cite{Casimir:1948dh, Bordag:1995gm, Bordag:2001qi}, the $L$ dependence of the total zero-point energy induces a pressure between parallel plates, a distance $L$ apart. At first the Casimir force was introduced as an attractive force between metallic plates. T.~H.~Boyer has found a repulsive force between the perfectly conducting and perfectly permeable plates \cite{Boyer:1974}. The connection of the attractive and repulsive forces has been studied for a perfect electromagnetic conductor in \cite{Rode:2017yqy}. Some useful formulae to calculate the Casimir force has been developed in arbitrary dimensions in \cite{Zhai:2014jta}. In a four-fermion interaction model the sign-flip phenomenon has been found in \cite{Flachi:2017cdo}.

In the four-fermion interaction model on $\mathcal{M}^{D-1} \otimes S^1$ the Casimir force, $F(L, \delta)$, is given as a function of $L$ and $\delta$. It is derived as the first derivative of the effective potential with respect to $L$ at the minimum, $\sigma = m$,
\begin{align}
 F(L,\delta) =& - \left. \frac{\partial \tilde{V}(\sigma)}{\partial L} \right|_{\sigma=m} .
 \label{def:casimir}
\end{align}
As is observed in Fig.~\ref{Minimum_Length}, the slope is negative (positive) for a small (large) $\delta$ and the repulsive (attractive) force induces. 

Substituting equation \eqref{EFSumInt} into \eqref{def:casimir}, we obtain the Casimir force at the leading order of the $1/N$ expansion.
\begin{align}
 \frac{F(L,\delta)}{m_\alpha^{D+1}}
 =& -\frac{\tr I}{(2\sqrt{\pi})^{D-1}\Gamma\left(\frac{D-1}{2}\right)}\frac{1}{(L m_\alpha)^2} \int_0^\infty \frac{\dmeasure{}{K}}{m_\alpha} \left(\frac{K}{m_\alpha}\right)^{D-2}\nonumber\\
 & \hspace{18ex}\times \left[ \ln \left(2\frac{\cosh \left(L\sqrt{K^2+m^2}\right) -\cos\left(\pi\delta\right)}{\exp\left( L\sqrt{K^2+m^2}\right)} \right)
 \right. \nonumber\\
 & \hspace{22ex}\left.
 + L\sqrt{K^2+m^2} \frac{\exp\left(-L\sqrt{K^2+m^2}\right) - \cos\left(\pi\delta\right)}{\cosh\left( L\sqrt{K^2+m^2}\right) - \cos\left(\pi\delta \right) }\right] ,
 \label{casimirForce:int}
\end{align}
where we set $\alpha=0$ and $1$ for $\lambda_r > \lambda_{cr}$ and $\lambda_r < \lambda_{cr}$, respectively. The fermion mass, $m$, is derived by solving the gap equation \eqref{gap:int}.

\begin{figure}[tbp]
 \begin{center}
  \begin{tabular}{cc}
   \begin{minipage}{0.33\hsize}
    \begin{center}
     \includegraphics[width=1\hsize]{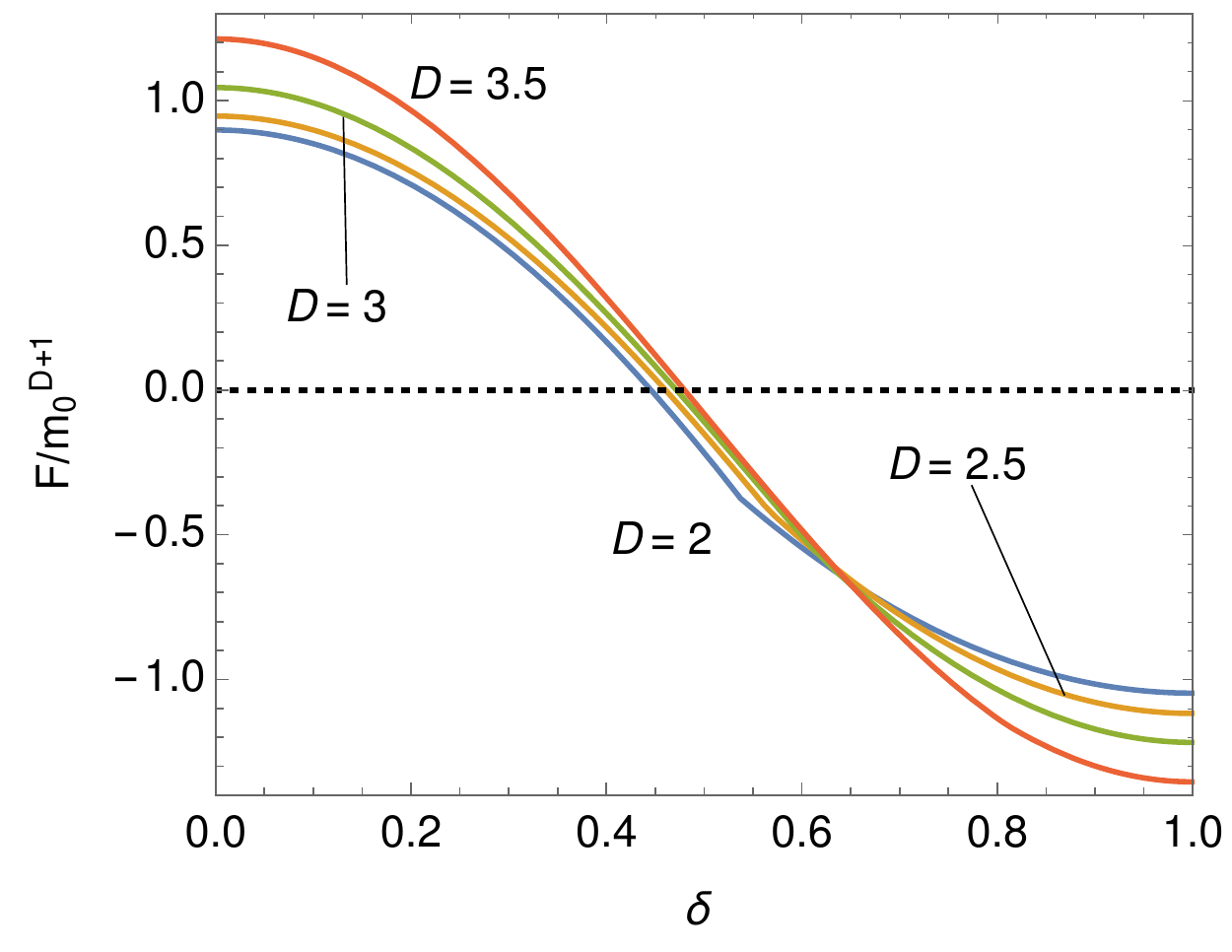}
     {(a) $\lambda_r>\lambda_{cr}$, $Lm_0=1$}
    \end{center}
    \vglue 2mm
   \end{minipage}
   &
   \begin{minipage}{0.33\hsize}
    \begin{center}
     \includegraphics[width=1\hsize]{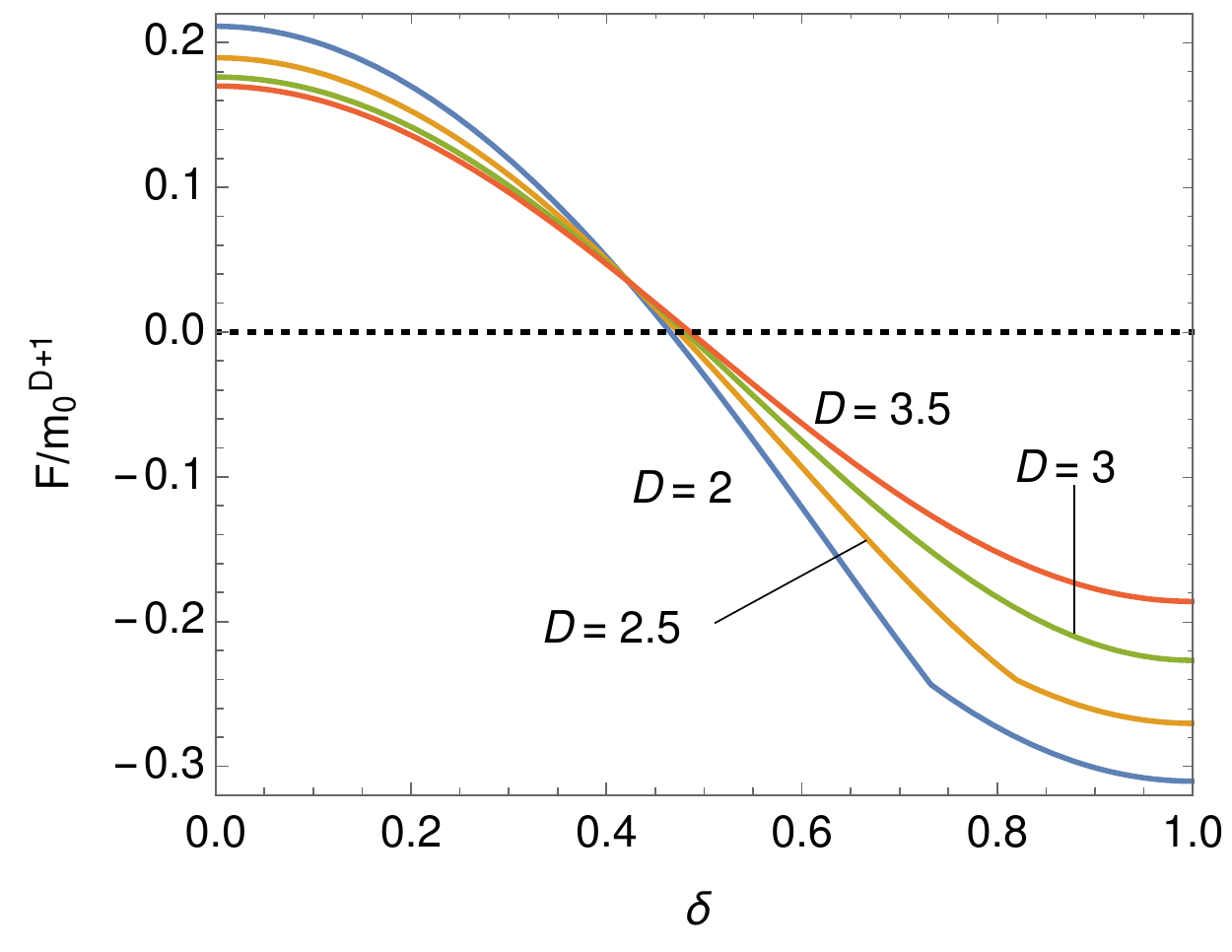}
     {(b) $\lambda_r>\lambda_{cr}$, $Lm_0=1.5$}
    \end{center}
    \vglue 2mm
   \end{minipage}
\\
   \begin{minipage}{0.33\hsize}
    \begin{center}
     \includegraphics[width=1\hsize]{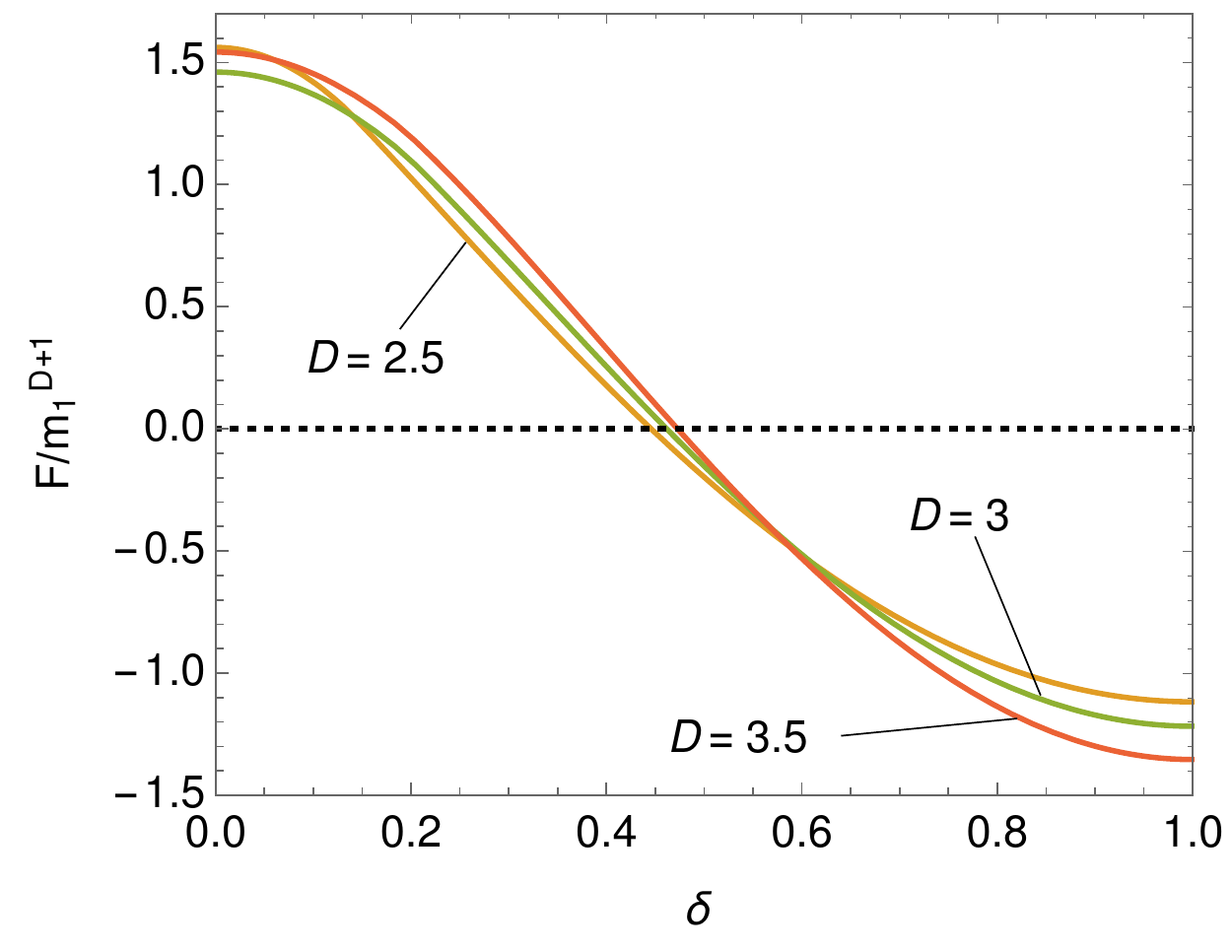}
     {(c) $\lambda_r<\lambda_{cr}$, $Lm_1=1$}
    \end{center}
   \end{minipage}
   &
    \begin{minipage}{0.33\hsize}
     \begin{center}
      \includegraphics[width=1\hsize]{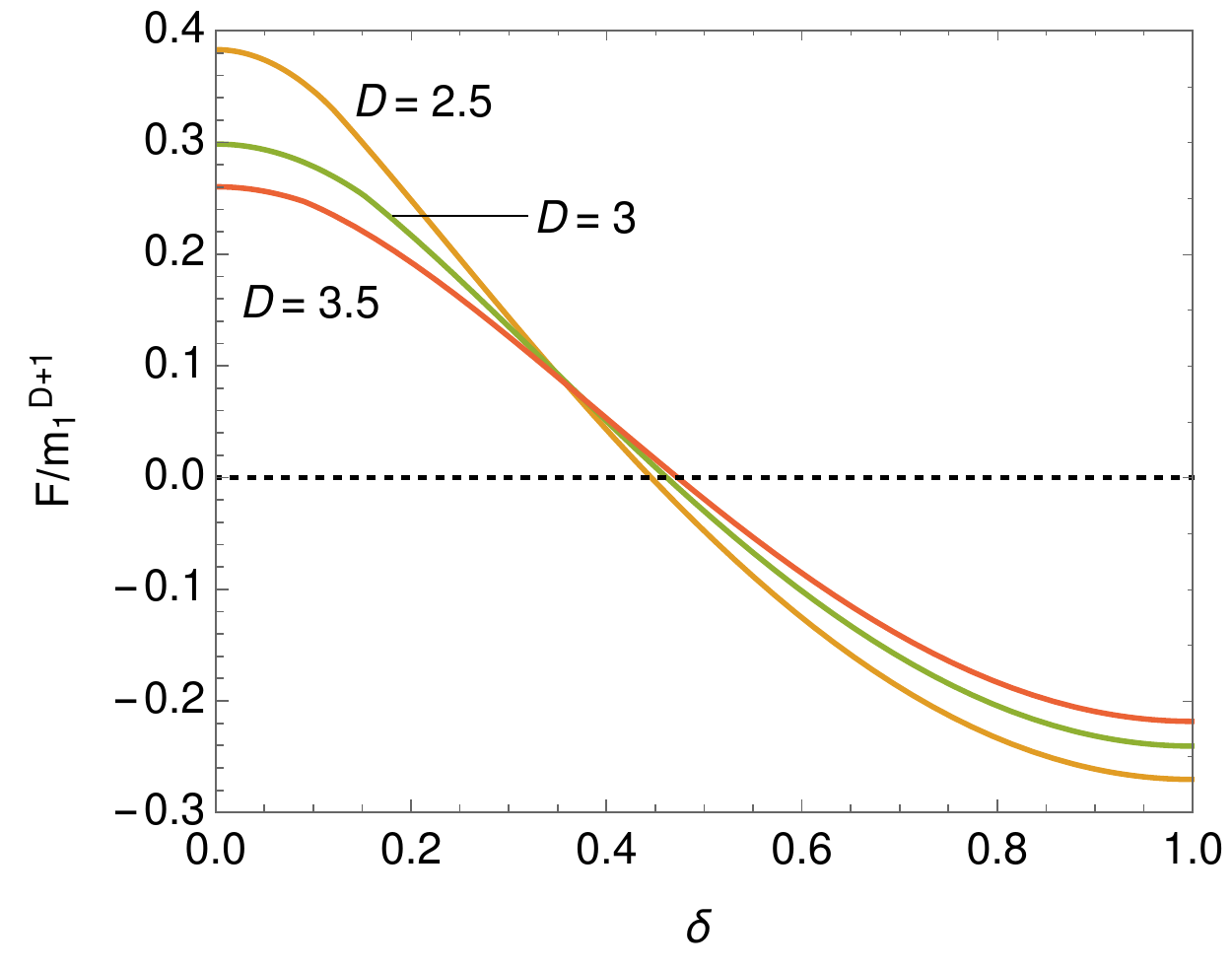}
      {(d) $\lambda_r<\lambda_{cr}$, $Lm_1=1.5$}
     \end{center}
    \end{minipage}
   \end{tabular}
  \caption{Casimir force as the function of $\delta$.}
  \label{CasimirForceDelta}
 \end{center}
\end{figure}

\begin{figure}[htbp]
 \begin{center}
  \begin{tabular}{ccc}
   \begin{minipage}{0.3\hsize}
    \begin{center}
     \includegraphics[width=1\hsize]{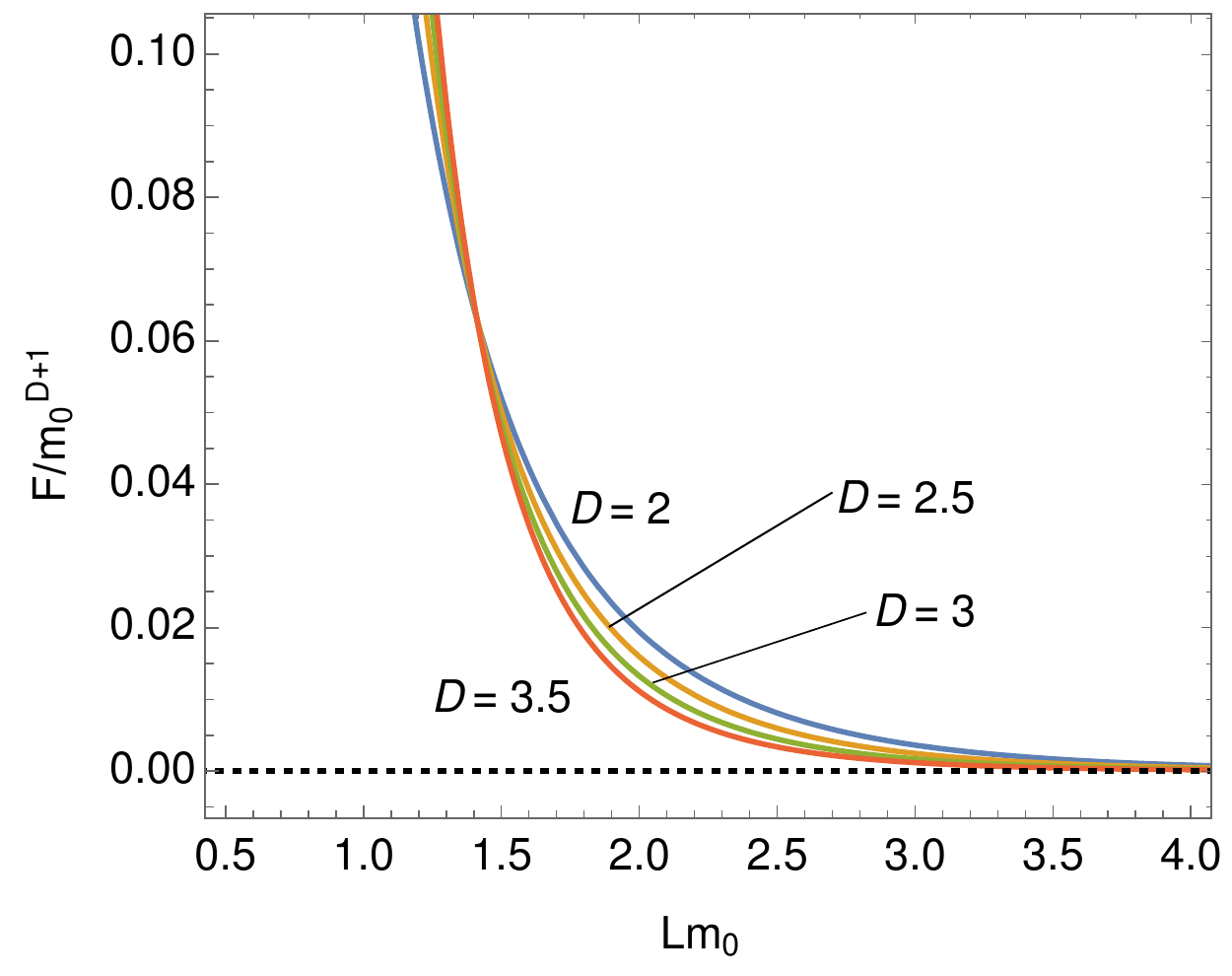}
     {(a) $\lambda_r>\lambda_{cr}$, $\delta = 0.4$}
    \end{center}
    \vglue 2mm
   \end{minipage}
   &
   \begin{minipage}{0.3\hsize}
    \begin{center}
     \includegraphics[width=1\hsize]{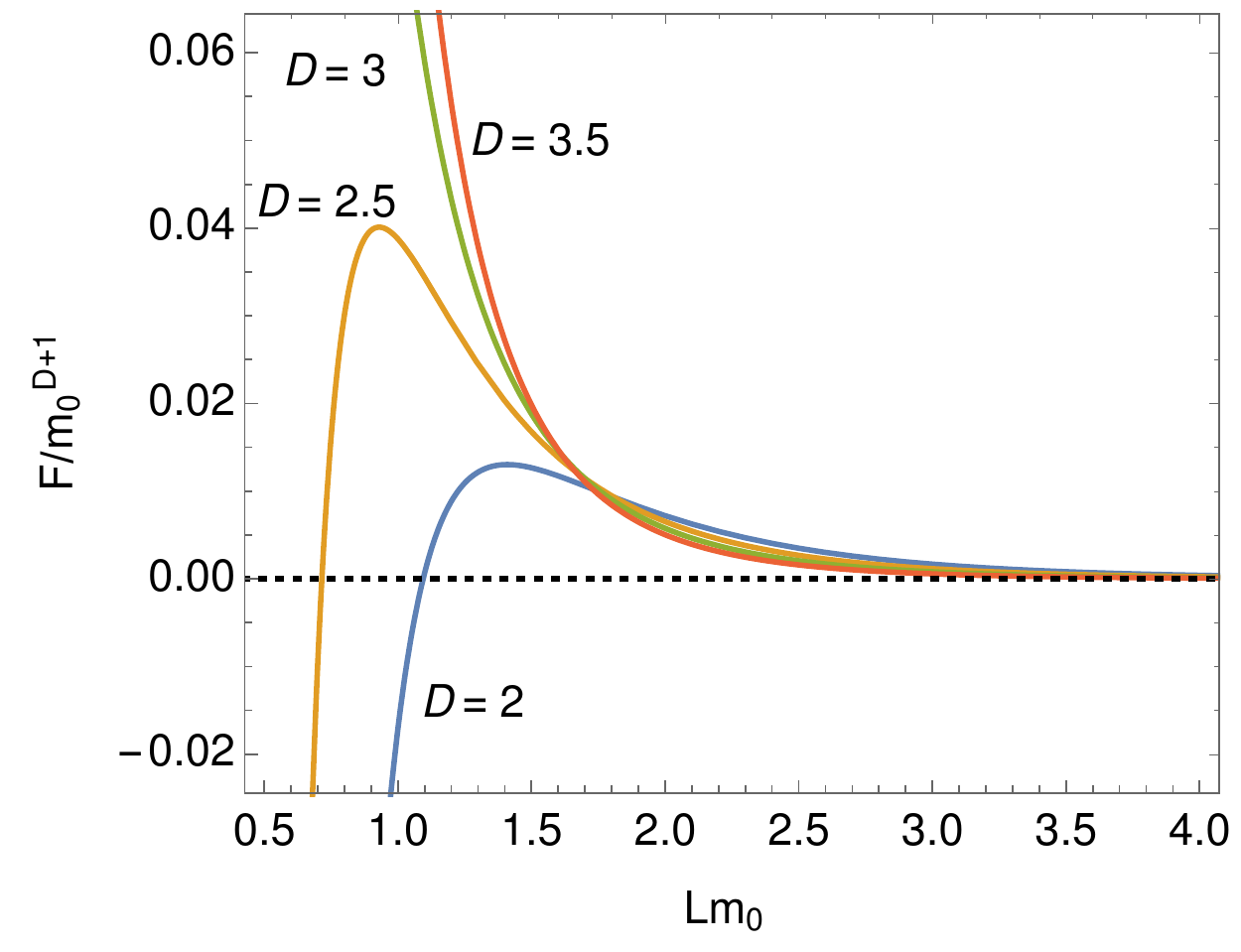}
     {(b) $\lambda_r>\lambda_{cr}$, $\delta = 0.45$}
    \end{center}
    \vglue 2mm
   \end{minipage}
   &
   \begin{minipage}{0.3\hsize}
    \begin{center}
     \includegraphics[width=1\hsize]{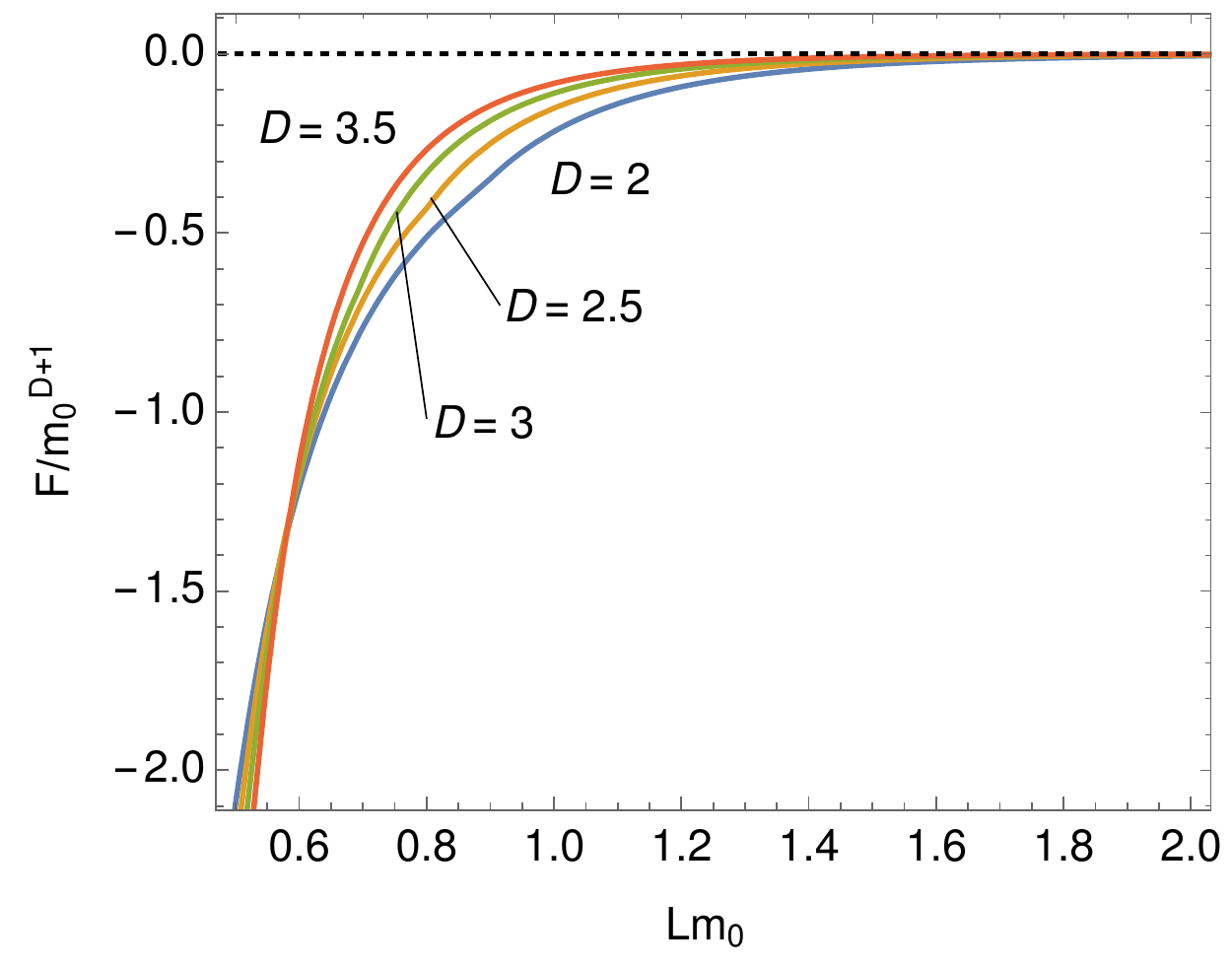}
     {(c) $\lambda_r>\lambda_{cr}$, $\delta = 0.5$}
    \end{center}
    \vglue 2mm
   \end{minipage}
\\
   \begin{minipage}{0.3\hsize}
    \begin{center}
     \includegraphics[width=1\hsize]
     {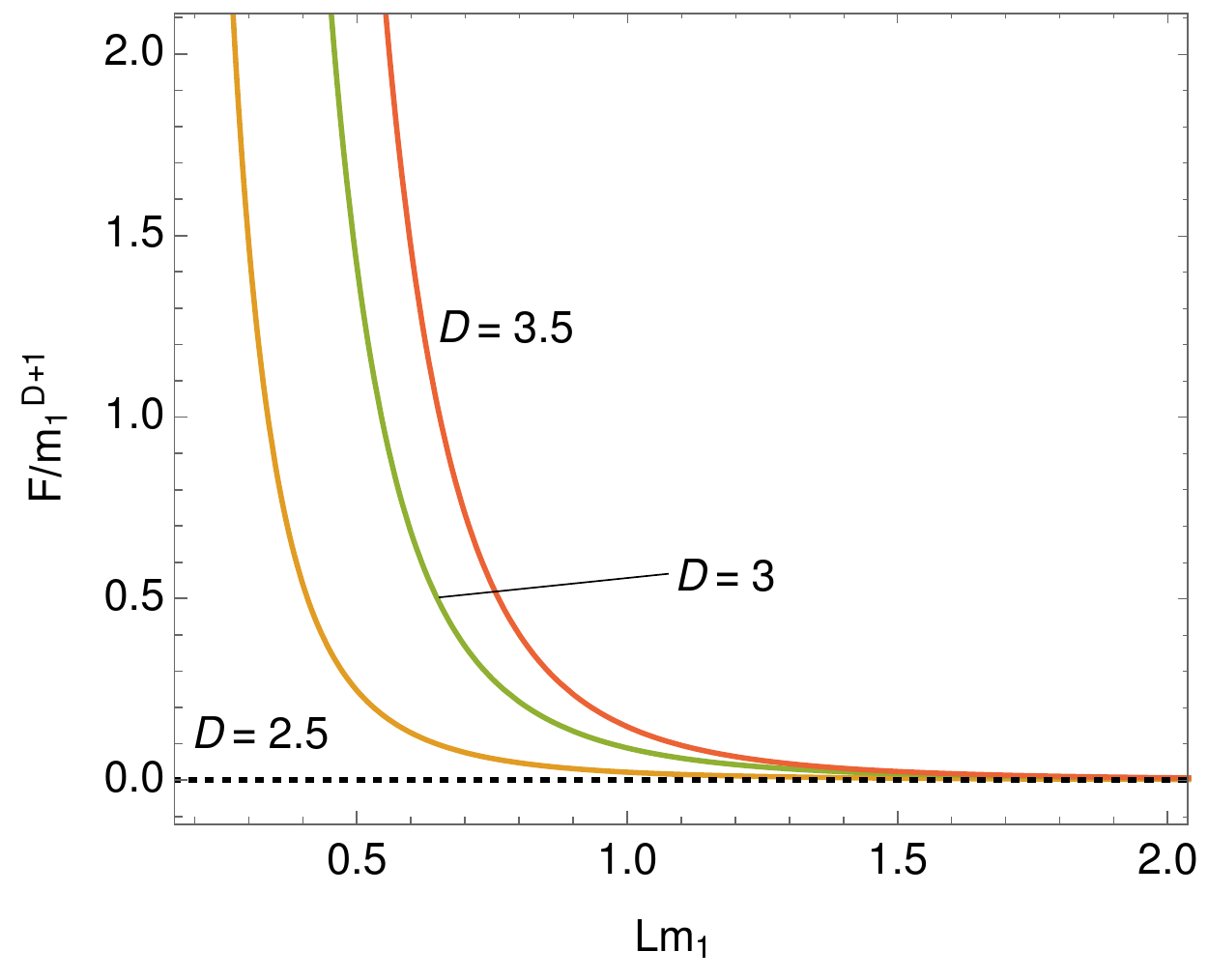}
     {(d) $\lambda_r<\lambda_{cr}$, $\delta = 0.44$}
    \end{center}
   \end{minipage}
   &
       \begin{minipage}{0.3\hsize}
	\begin{center}
	 \includegraphics[width=1\hsize]
	 {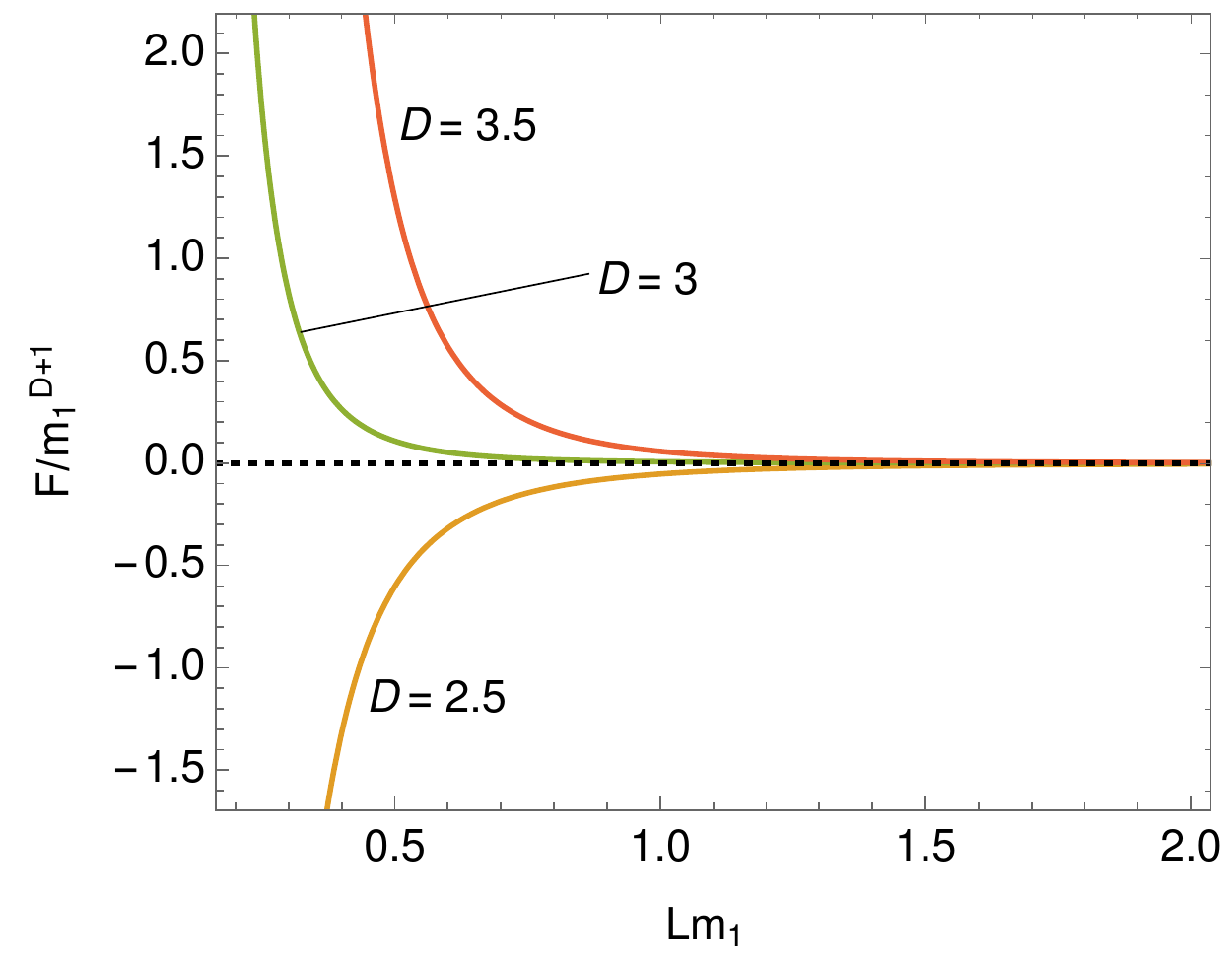}
	 {(e) $\lambda_r<\lambda_{cr}$, $\delta = 0.46$}
	\end{center}
       \end{minipage}
       &
	   \begin{minipage}{0.3\hsize}
	    \begin{center}
	     \includegraphics[width=1\hsize]
	     {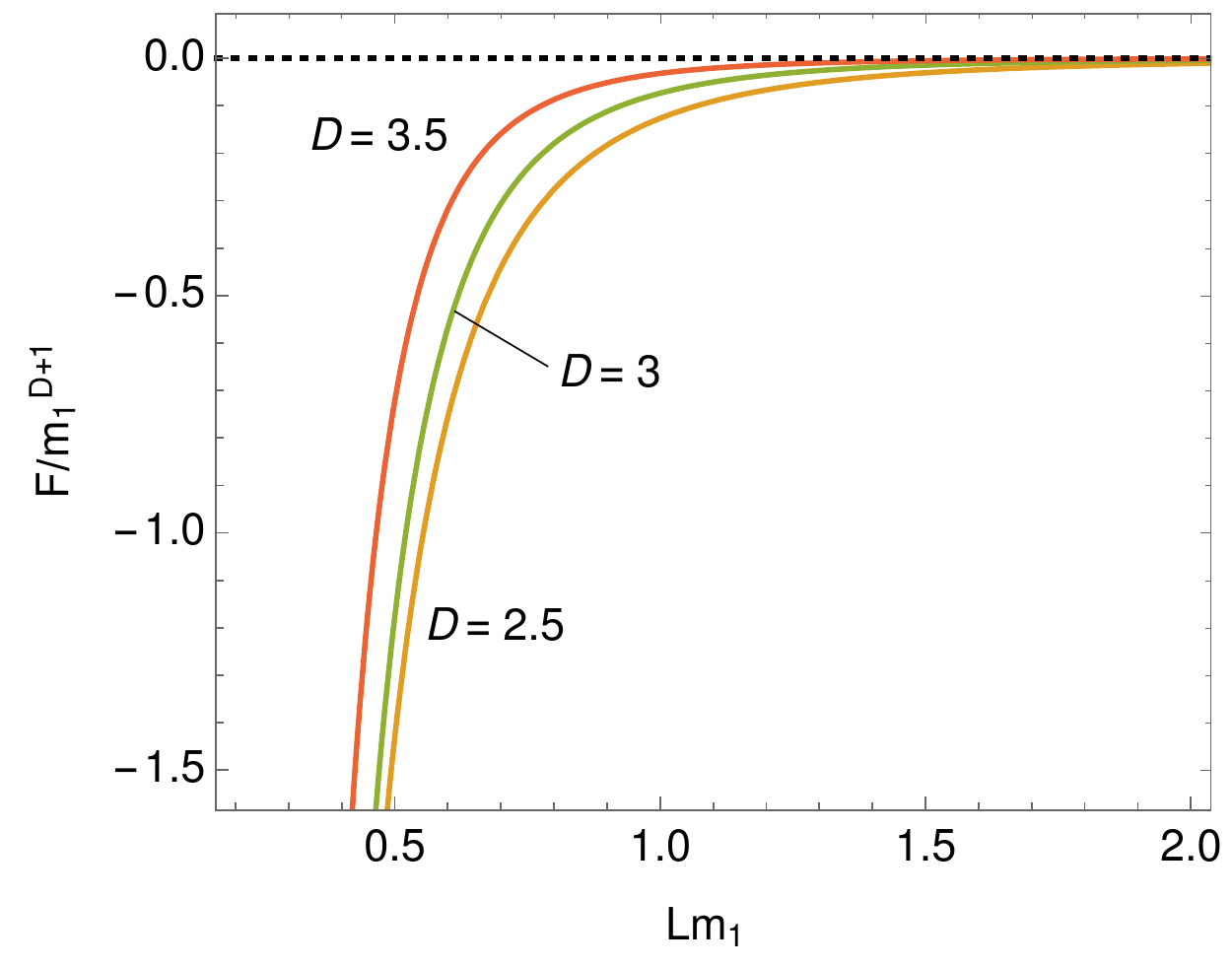}
	     {(f) $\lambda_r<\lambda_{cr}$, $\delta = 0.48$}
	    \end{center}
	   \end{minipage}
  \end{tabular}
  \caption{Casimir force as a function of $L$.}
  \label{CasimirForceLength}
 \end{center}
\end{figure}

In Fig. \ref{CasimirForceDelta} we plot the behavior of the Casimir force as a function of the phase $\delta$ for fixed lengths. As typical lengths, we choose $Lm_\alpha = 1$ and $1.5$. The sharp bends on the lines, it is clearly observed in two dimensions, $D=2$, for $\lambda > \lambda_{cr}$, correspond to the critical value for the chiral symmetry breaking. It is shown that the Casimir force is repulsive near the periodic boundary condition, $\delta=0$, and the force decreases monotonically as approaching the anti-periodic boundary condition, $\delta=1$. The sign of $F(L, \delta)$ changes around $\delta \sim 0.5$ for $D=2, 2.5, 3$ and $3.5$. The attractive force realizes near the anti-periodic boundary condition.

In Fig. \ref{CasimirForceLength}  the behavior of the Casimir force is plotted as a function of $L$ around the sign-flip value, $\delta \sim 0.5$. We observe that the force is divergent and disappears at the limit $L\to 0$ and $L\to\infty$, respectively. The sign flips in two cases $D=2, 2.5$ at  $\delta=0.45$ for a strong coupling, $\lambda > \lambda_{cr}$. In these cases the Casimir force changes from attractive to repulsive as the length increases. Thus the length where the sign flips is unstable. No stable length is realized for the four-fermion interaction model. In the other cases the sign-flip is not observed. 
 
The sign-flip points for the Casimir force are found by solving $F(L,\delta)=0$. In the symmetric phase, $m=0$, the momentum integral in \eqref{casimirForce:int} is described by the polylogarithm, $\mathrm{Li}_s(z)$, which is defined in \eqref{polylogarithm}. Then equation \eqref{casimirForce:int} reduces to
 \begin{align}
  \frac{F(L,\delta)_{\mathrm{sym}}}{m_\alpha^{D+1}}
  &= \frac{\tr I \cdot \Gamma\left(D+1\right)}{(2\sqrt{\pi})^{D-1} \Gamma\left(\frac{D+1}{2}\right)} \frac{1}{(Lm_\alpha)^{D+1}} \mathrm{Re}\,\mathrm{Li}_D\left(e^{i\pi\delta}\right). \label{CasimirForceSym}
 \end{align}
 Thus the sign-flip points in the symmetric phase are found to be
 \begin{align}
  \mathrm{Re}\, \mathrm{Li}_D(e^{i\pi\delta}) = 0 .
 \end{align}
Since this equation is independent of the length, $L$, the $\mathrm{U}(1)$ phase, $\delta$, fixes the sign-flip points. We give the phase $\delta$ at the sign-flip points for $D=2, 2.5, 3$ and $3.5$ in Tab.~\ref{sign-flip:delta}. In the broken phase we numerically evaluate equation \eqref{casimirForce:int} with the gap equation \eqref{gap:int} and find the solution for $F(L,\delta)=0$.

\begin{table}[htbp]
\begin{center}
  \begin{tabular}{l|cccc}
    $D$ & 2 & 2.5 & 3 & 3.5 \\
    \hline
     $\delta$ & 0.42265 & 0.44575 & 0.46166 & 0.47280
  \end{tabular}
  \end{center}
  \caption{Phase $\delta$ at the sign-flip points in the symmetric phase.}
  \label{sign-flip:delta}
\end{table}

\begin{figure}[tbp]
 \begin{center}
  \begin{tabular}{cc}
   \begin{minipage}{0.4\hsize}
    \begin{center}
     \includegraphics[width=1\hsize]{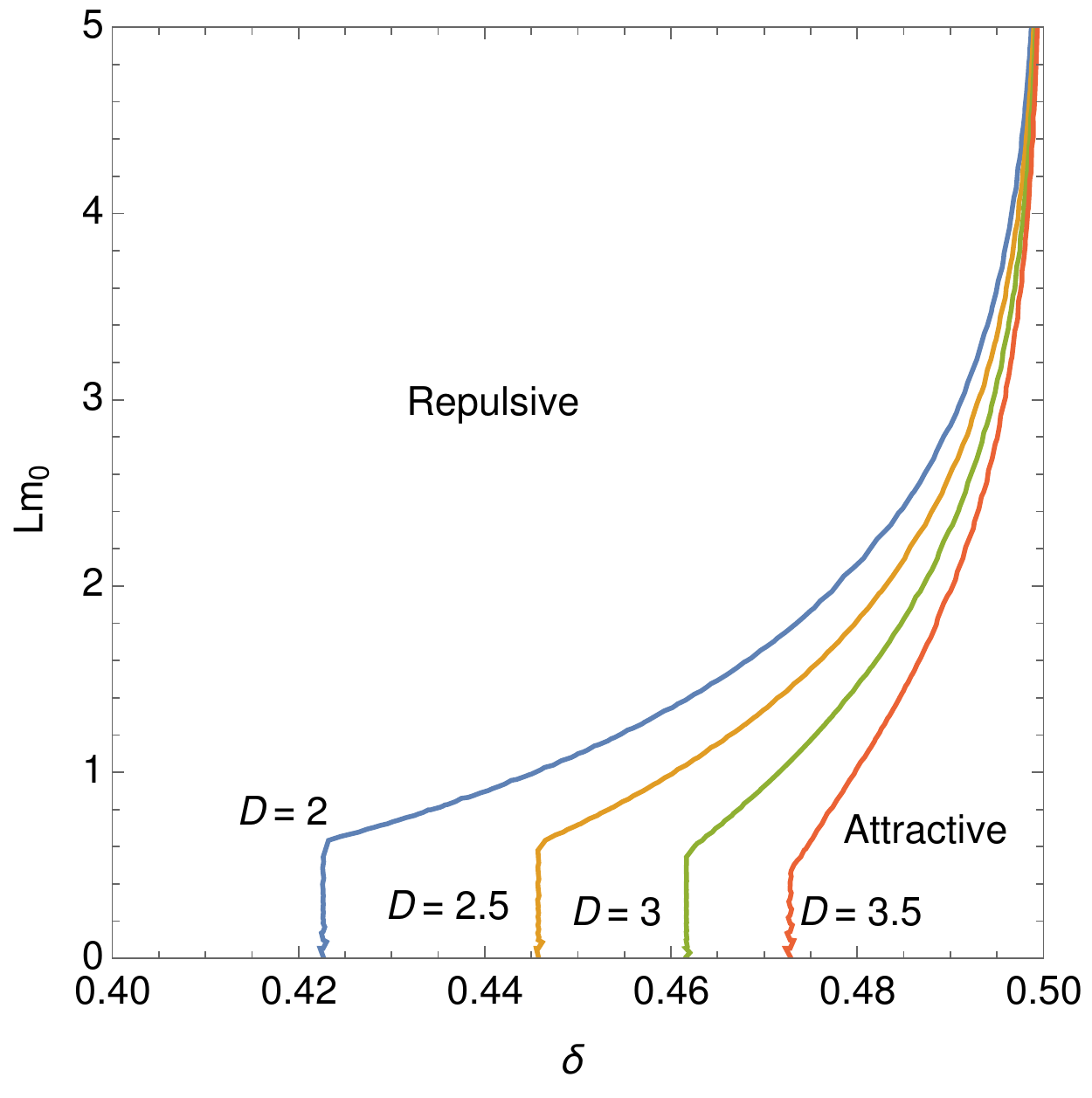}
     {(a) $\lambda_r>\lambda_{cr}$}
    \end{center}
   \end{minipage}
   &
   \begin{minipage}{0.4\hsize}
    \begin{center}
     \includegraphics[width=1\hsize]{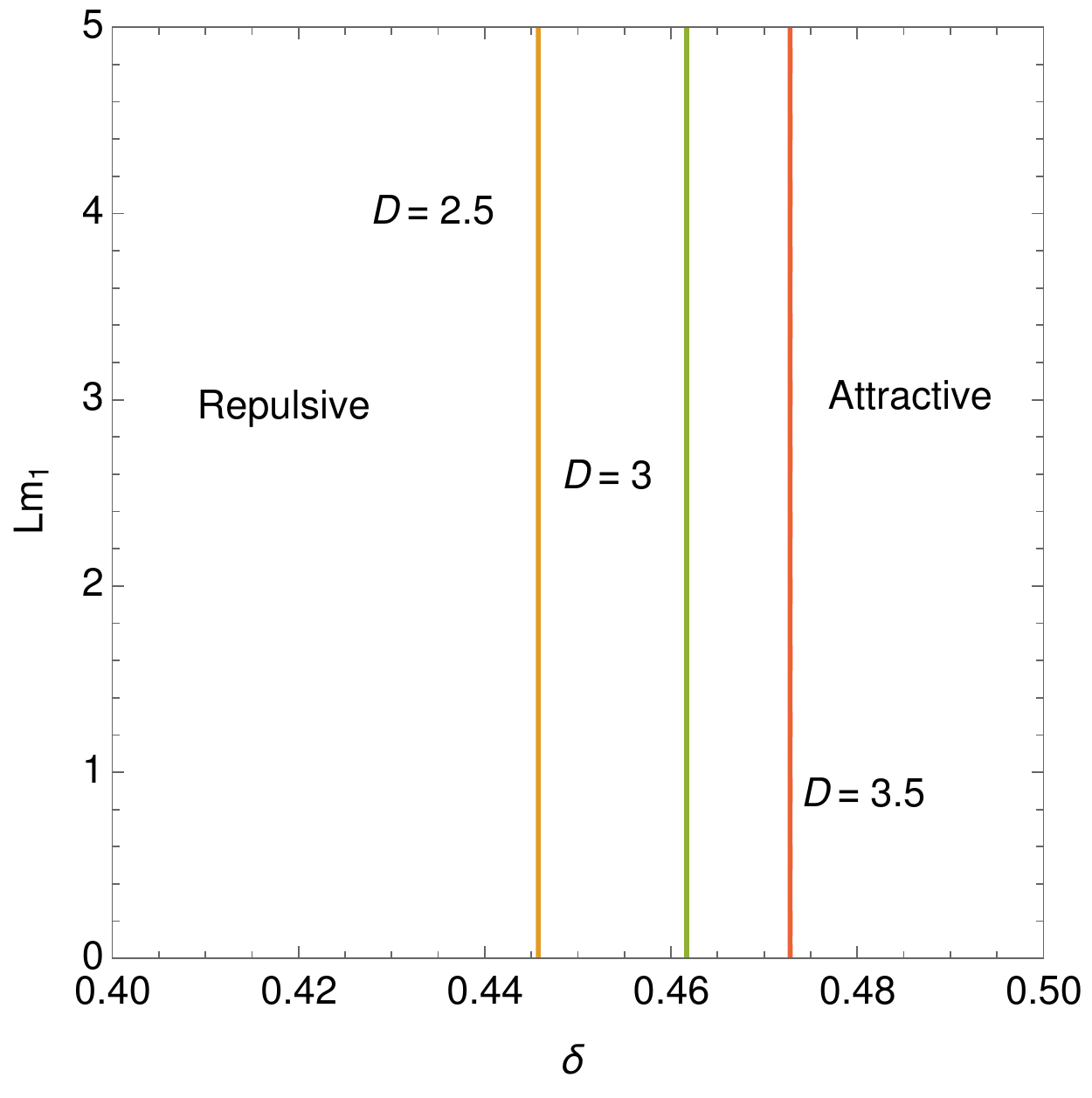}
     {(b) $\lambda_r<\lambda_{cr}$}
    \end{center}
   \end{minipage}
  \end{tabular}
  \caption{Boundary between the repulsive (left side of the lines) and attractive (right side of the lines) force.}
  \label{CasimirForceZeroPoints}
 \end{center}
\end{figure}

In Fig.~\ref{CasimirForceZeroPoints} we draw the boundary lines dividing the repulsive force and attractive force on the $L-\delta$ plane. In the symmetric phase the boundary is fixed by the phase, $\delta$, and described by a vertical lines. As is shown in Fig.~\ref{CriticalLength}, the chiral symmetry is restored when the length, $L$, is smaller than the critical one for $\lambda_r >\lambda_{cr}$. In Fig.~\ref{CasimirForceZeroPoints} (a), we observe that the boundary is represented by vertical line in the symmetric phase. Above the critical length, the length at the boundary increases as $\delta$ approaches $0.5$. The dynamically generated fermion mass extends the domain where the repulsive force is induced. In Fig.~\ref{CasimirForceZeroPoints} (b), the sign-flip points are found on the symmetric phase and the boundary is represented by vertical lines. 

 \section{Conclusions}
 
We have studied dynamical chiral symmetry breaking in four-fermion interaction models on $\mathcal{M}^{D-1} \otimes S^1$ with the $\mathrm{U}(1)$-valued boundary condition. The models are often considered associated with a superconducting ring, non-trivial topology at the early universe and compact extra dimensions. Assuming the homogeneous condensation and using the zeta function regularization, we have obtained the explicit expression of the effective potential for the fermion and anti-fermion composite field in the leading order of the $1/N$ expansion.  

The system is classified into two cases based on the chiral symmetry on $\mathcal{M}^{D}$. In the strong coupling case, $\lambda_r > \lambda_{cr}$, the composite field develops a non-vanishing expectation value and the chiral symmetry is dynamically broken, while the expectation value for the composite field vanishes and the ground state maintains chiral symmetry in the weak coupling case, $\lambda_r > \lambda_{cr}$, on $\mathcal{M}^{D}$. In the specific expressions these cases are distinguished by the mass scale, $m_0$ and $m_1$. No constraints is theoretically defined for the mass scale and it is fixed for each phenomenon.

In this paper we focus on the topological effect stemming from the boundary condition. The effective potential was numerically evaluated on $\mathcal{M}^{D-1} \otimes S^1$ as the $\mathrm{U}(1)$ phase, $\delta$, varies. By observing the effective potential at the minimum, the stable state with respect to the $\mathrm{U}(1)$ phase is found at the anti-periodic boundary condition, $\delta=1$. We calculated the dynamically generated fermion mass as a function of $\delta$ and checked that only the second order phase transition takes place. The phase diagram was shown on the $\delta-L$ and $D-L$ planes for $\lambda_r > \lambda_{cr}$ and $\lambda_r < \lambda_{cr}$ in Fig.~\ref{CriticalLength}.

To find a phenomenological consequence the Casimir force has been investigated in the models on $\mathcal{M}^{D-1} \otimes S^1$. As is pointed out in \cite{Flachi:2017cdo}, the sign of the force flips as $\delta$ varies from $0$ to $1$. We found the explicit expression for the sign-flip points in the symmetric phase and the boundary lines dividing the repulsive force and attractive force on the $L-\delta$ plane. In Fig.~\ref{CasimirForceZeroPoints} (a)  the critical points for chiral symmetry breaking are clearly observed as sharp bends on the lines.

The derived expressions for the effective potential reduce to the known results in the previous works at the periodic ($\delta=0$) and anti-periodic ($\delta=1$) boundary conditions \cite{Inagaki:1997kz,Kim:1994es, Ishikawa:1998uu}. It should be noted that the imaginary chemical potential introduces similar expressions \cite{Matsumoto:2010vw}.

In the present work we assume the homogeneous expectation value for the composite field and study the ground state by calculating the effective potential. In finite size space-times, inhomogeneous states may be realized. The inhomogeneous states are found by observing the effective action on $\mathcal{M}^{D-1} \otimes S^1$ or extend the analysis developed in two dimensions \cite{Yoshii:2014fwa}. Four-fermion interaction models in a curved geometory with a non-trivial topology are also interesting to find some phenomenological consequences at the early universe \cite{Inagaki:1995jp}. We hope to report on the inhomogeneous condensation in four-fermion interaction models on $\mathcal{M}^{D-1} \otimes S^1$ and a curved geometory in future.

 \section*{Acknowledgements}
 The authors would like to thank R.~Yoshii, K.~Ishikawa and H.~Sakamoto for valuable discussions. Discussions during the Riken symposium on ``Thermal Quantum Field Theories and Their Applications 2018'' were useful to complete this work.

 \appendix
 \section*{Appendix: Effective potential on $\mathcal{M}^{D-1} \otimes S^1$}\label{AppendixA}

Here we present details of the calculation of the effective potential \eqref{EFwithFinite} and show two types of expressions.
The integrand in the second term of the effective potential is calculated as
\begin{align}
   \tr\ln \frac{ \gamma^\mu K_\mu - \omega_n \gamma^{D-1}- \sigma }{-\omega}
  &= \tr\ln \frac{ \sigma }{\omega}+ \tr\ln \left(1- \frac{\gamma^\mu K_\mu - \omega_n \gamma^{D-1}}{\sigma}\right)
  \nonumber \\
  &= \tr\ln \frac{ \sigma }{\omega} - \tr\sum_{k=1}^{\infty} \frac{1}{k}\left( \frac{\gamma^\mu K_\mu - \omega_n \gamma^{D-1}}{\sigma}\right)^k
 \nonumber \\
  &= \tr I \ln \frac{ \sigma }{\omega} - \tr I \sum_{k=1}^{\infty} \frac{1}{2k}\left( \frac{K^2 - \omega_{n}^2 }{\sigma ^2 }\right)^k
  \nonumber \\
  &= \frac{\tr I }{2} \ln \frac{- K^2 + \omega_{n}^2 +\sigma^2 }{\omega^2}.
\end{align}
In going from the second to third line, we trace over the Dirac indices. After the Wick rotation, $K^0 \to i K^0$, the second term of \eqref{EFwithFinite} reads
 \begin{align}
  I_{2nd}=\frac{i}{L}\sum_{n=-\infty}^{\infty} \int\frac{\dmeasure{D-1}{K}}{(2\pi)^{D-1}} \tr\ln \frac{ \gamma^\mu K_\mu - \omega_n \gamma^{D-1}- \sigma }{-\omega}
  &= -\frac{\tr I}{2L}\sum_{n=-\infty}^{\infty} \int\frac{\dmeasure{D-1}{K}}{(2\pi)^{D-1}} \ln\frac{ K^2 + \omega_n^2 + \sigma^2 }{\omega^2}. \label{Appendix:EFFiniteSqure}
 \end{align}
Using the following formula which comes from the zeta function regularization,
\begin{align}
  \tr \ln \mathcal{O} = - \lim_{s \to 0} \frac{\diff}{\diff s} \frac{1}{\Gamma\left(s \right)}\int_0^\infty \dmeasure{}{t} t^{s-1} \tr e^{-t\mathcal{O}},
\end{align}
and employing the formula,
 \begin{align}
  \sum_{n=-\infty}^\infty e^{-\frac{i\pi}{\tau}(n+z)^2} = \sqrt{\frac{\tau}{i}}\vartheta_3\left(z |\tau\right),
\end{align}
with the definition for the Jacobi theta function, $\vartheta_3(z |\tau)$,
\begin{align} 
  \vartheta_3\left(z | \tau\right)
  = \sum_{n=-\infty}^\infty e^{\pi i \tau n^2 + 2\pi iz n}
  = 1 + 2\sum_{n=1}^\infty \left(e^{\pi i \tau} \right)^{n^2} \cos (2n\pi z) ,
 \end{align}
then equation \eqref{Appendix:EFFiniteSqure} reads
 \begin{align}
  I_{2nd}= &\lim_{s \to 0} \frac{\tr I}{2L} \int\frac{\dmeasure{D-1}{K}}{(2\pi)^{D-1}} \frac{\diff}{\diff s} \frac{1}{\Gamma(s)} \int_0^\infty \dmeasure{}{t} t^{s-1} e^{-t\frac{K^2 + \sigma^2}{\omega^2}} \sum_{-\infty}^{\infty} e^{-\frac{4\pi^2t}{L^2\omega^2}\left(n+\frac{\delta}{2}\right)^2 }  
  \nonumber \\
 &=\lim_{s \to 0} \frac{\tr I \omega}{4\sqrt{\pi}} \int\frac{\dmeasure{D-1}{K}}{(2\pi)^{D-1}}
  \frac{\diff}{\diff s} \frac{1}{\Gamma(s)}  \left( \int_0^\infty \dmeasure{}{t} t^{s-\frac{3}{2}} e^{-t\frac{K^2 + \sigma^2}{\omega^2}} \right.
  \nonumber \\
  & \hspace{30ex} \left .+ 2 \sum_{n=1}^\infty \cos (n\pi\delta) \int_0^\infty \dmeasure{}{t} t^{s-\frac{3}{2}} e^{-\frac{L^2\omega^2 n^2 }{4t} -t\frac{K^2 + \sigma^2}{\omega^2}} \right) .
\label{Appendix:EFFiniteSqure:Zeta}
 \end{align}
The $t$-integrations in \eqref{Appendix:EFFiniteSqure:Zeta} are represented by the gamma function and the modified Bessel function of the second kind,
\begin{align}
  &\int_0^\infty \dmeasure{}{t} t^{s-\frac{3}{2}} e^{-t\frac{K^2 + \sigma^2}{\omega^2}}
  = \left(\frac{K^2 + \sigma^2}{\omega^2}\right)^{-s + \frac{1}{2}} \Gamma\left(s-\frac{1}{2}\right),\label{Appendix:int1}\\
  &\int_0^\infty \dmeasure{}{t} t^{s-\frac{3}{2}} e^{-\frac{L^2\omega^2 n^2 }{4t} -t\frac{K^2 + \sigma^2}{\omega^2} }
  = 2^{\frac{3}{2}-s} \left( \frac{K^2 + \sigma^2}{L^2 \omega^4 n^2}\right)^{\frac{1}{4} - \frac{s}{2}} \mathrm{K}_{\frac{1}{2}-s}\left(Ln\sqrt{K^2+\sigma^2}\right).\label{Appendix:int2}
 \end{align}
Substituting the equations \eqref{Appendix:int1}, \eqref{Appendix:int2} into \eqref{Appendix:EFFiniteSqure:Zeta} and taking the $s \to 0$ limit, we obtain
  \begin{align}
   I_{2nd}= -\frac{\tr I}{2} \int\frac{\dmeasure{D-1}{K}}{(2\pi)^{D-1}}
   \left[ \left(K^2 + \sigma^2\right)^{\frac{1}{2}}
   - \frac{2\sqrt{2}}{\sqrt{\pi L}}\left( K^2 + \sigma^2\right)^{\frac{1}{4}} \sum_{n=1}^\infty n^{-\frac{1}{2}} \mathrm{K}_{\frac{1}{2}}\left(Ln\sqrt{K^2+\sigma^2}\right) \cos (n\pi\delta) \right].
  \end{align}
Thus the effective potential \eqref{EFwithFinite} reads
  \begin{align}
   V(\sigma)=& \frac{1}{2\lambda_0}\sigma^2-\frac{\tr I}{2} \int\frac{\dmeasure{D-1}{K}}{(2\pi)^{D-1}}
   \left[ \left(K^2 + \sigma^2\right)^{\frac{1}{2}} \right. \nonumber \\
   &\left.\hspace{6ex} - \frac{2\sqrt{2}}{\sqrt{\pi L}}\left( K^2 + \sigma^2\right)^{\frac{1}{4}} \sum_{n=1}^\infty n^{-\frac{1}{2}} \mathrm{K}_{\frac{1}{2}}\left(Ln\sqrt{K^2+\sigma^2}\right) \cos (n\pi\delta) \right].
   \label{Appendix:epot}
  \end{align}
The effective potential \eqref{epot0} is also calculated along the same procedure. It is easy to find by taking the $L \to \infty$ limit of \eqref{Appendix:epot},
\begin{align}
   V_0(\sigma)=\lim_{L\to\infty}V(\sigma)=& \frac{1}{2\lambda_0}\sigma^2-\frac{\tr I}{2} \int\frac{\dmeasure{D-1}{K}}{(2\pi)^{D-1}}
    \left(K^2 + \sigma^2\right)^{\frac{1}{2}} .
   \label{Appendix:epot0}
\end{align}
The summation in \eqref{Appendix:epot} is performed by using the formula,
  \begin{align}
   \sum_{n=1}^\infty n^{-\frac{1}{2}} \mathrm{K}_{\frac{1}{2}}\left(L\sqrt{K^2 + \sigma^2} n\right) \cos \left(n\pi\delta\right)
   = -\frac{1}{2\sqrt{L}\left(K^2 + \sigma^2\right)^\frac{1}{4}} \sqrt{\frac{\pi}{2}} \ln \left( 2\frac{\cosh \left(L\sqrt{K^2 + \sigma^2}\right) - \cos \left(\pi\delta \right)}{\exp\left( L\sqrt{K^2 + \sigma^2}\right) } \right).
  \end{align}
After the angular integration we obtain
  \begin{align}
  V(\sigma) &= V_0(\sigma)
  - \frac{\tr I}{\left(2\sqrt{\pi}\right)^{D-1} \Gamma\left(\frac{D-1}{2}\right)} \frac{1}{L}\int_0^\infty \dmeasure{} K K^{D-2} \ln \left(2\frac{\cosh\left(L\sqrt{K^2+\sigma^2} \right) -\cos\left(\pi\delta\right)}{\exp\left( L\sqrt{K^2+\sigma^2}\right)} \right) .
   \label{Appendix:epot:int}
  \end{align}
Since the second term of the right hand side in \eqref{Appendix:epot:int} is finite, the divergent zero-point energy is removed by subtracting $V_0(\sigma=0)$.
Therefore the expression of the effective potential \eqref{EFwithFiniteSumInt} is derived.

Next we rewrite equation \eqref{Appendix:EFFiniteSqure} as
 \begin{align}
  I_{2nd}= -\frac{\tr I}{2L}\sum_{n=-\infty}^{\infty} \int\frac{\dmeasure{D-1}{K}}{(2\pi)^{D-1}} \ln\frac{ K^2 + \omega_n^2 + \sigma^2 }{K^2}
 +C(L),
 \label{Appendix:EFFiniteSqure2}
 \end{align}
 with the $\sigma$ independent function, $C(L)$, comes from the zero-point energy,
\begin{align}
   C(L)
   =    -\frac{\tr I}{2L}\sum_{n=-\infty}^{\infty} \int\frac{\dmeasure{D-1}{K}}{(2\pi)^{D-1}} \ln\frac{ K^2}{\omega^2} .
   \label{Appendix:C}
\end{align}
Following the procedure developed in \cite{Inagaki:1994ec}, we perform the momentum integral of the first term in the right hand side of \eqref{Appendix:EFFiniteSqure2} and get
 \begin{align}
  I_{2nd}= \frac{\tr I }{2L(2\sqrt{\pi})^{D-1}}\Gamma\left(\frac{1-D}{2}\right)\sum_{n=-\infty}^{\infty} 
  (\omega_n^2+\sigma^2)^\frac{D-1}{2}
  +C(L).
 \label{Appendix:EFFiniteSqure3}
 \end{align}
Then we obtain
\begin{align}
  V(\sigma) =  \frac{1}{2\lambda_0}\sigma^2
   + \frac{\tr I }{2L(2\sqrt{\pi})^{D-1}}\Gamma\left(\frac{1-D}{2}\right)\sum_{n=-\infty}^{\infty} 
  (\omega_n^2+\sigma^2)^\frac{D-1}{2}
  +C(L) .
   \label{Appendix:epot2}
  \end{align}
The summation on the right hand side of \eqref{Appendix:epot2} is divergent. 
The divergence is regularized by using the following formula \cite{Elizalde:Zeta},
  \begin{align}
   \sum_{n=-\infty}^\infty \left[ a(n+c)^2 + q\right]^{-s}
   = \frac{\sqrt{\pi}}{a^{\frac{1}{2}}} \frac{\Gamma\left(s-\frac{1}{2}\right)}{\Gamma\left(s\right)} q^{\frac{1}{2}-s}
   + \frac{4\pi^s}{\Gamma\left(s\right)} \frac{\left(a q\right)^{\frac{1-2s}{4}}}{a^{\frac{1}{2}}} \sum_{n=1}^\infty \frac{\cos \left(2\pi c n\right)}{n^{\frac{1}{2} -s}} \mathrm{K}_{\frac{1}{2}-s}\left(2\pi n \left(\frac{q}{a}\right)^{\frac{1}{2}}\right) .
  \end{align}
Thus the effective potential reads
   \begin{align}
   V(\sigma)
   = \frac{1}{2\lambda_0}\sigma^2 + \frac{\tr I }{2(4\pi)^{D/2}}\Gamma\left(-\frac{D}{2}\right)(\sigma^2)^\frac{D}{2}
   + 2\tr I \left(\frac{\sigma}{2\pi L}\right)^{\frac{D}{2}} \sum_{n=1}^\infty \frac{\cos \left(n\pi \delta\right)}{n^{\frac{D}{2}}}\mathrm{K}_{\frac{D}{2}}\left(L\sigma n\right)
  +C(L).
   \label{EFwithFiniteIntSumConv0}
  \end{align}
The second term on the right hand side of \eqref{EFwithFiniteIntSumConv0} is divergent in even dimensions. The divergent term is equivalent to the one in \eqref{Appendix:epot0}. After the momentum integral equation \eqref{Appendix:epot0} reduces to
\begin{align}
   V_0(\sigma) = \frac{1}{2\lambda_0}\sigma^2 - \frac{\tr I }{(4\pi)^{D/2} D}\Gamma\left(1-\frac{D}{2}\right)(\sigma^2)^\frac{D}{2} +C(L\to\infty).
   \label{Appendix:epot0:2}
  \end{align}
Substituting equation \eqref{Appendix:epot0:2} into \eqref{EFwithFiniteIntSumConv0}, the regularized expression for the effective potential is derived as
\begin{align}
   V(\sigma)
   = V_0(\sigma)
   + 2\tr I \left(\frac{\sigma}{2\pi L}\right)^{\frac{D}{2}} \sum_{n=1}^\infty \frac{\cos\left( n\pi \delta\right)}{n^{\frac{D}{2}}}\mathrm{K}_{\frac{D}{2}}\left(L\sigma n\right) +C(L) -C(L\to\infty).
   \label{EFwithFiniteIntSumConv}
\end{align}

For $\sigma=0$ the effective potential \eqref{Appendix:epot2} is simplified to
 \begin{align}
  V(0)
  &= \frac{\tr I  }{2L (2\sqrt{\pi})^{D-1}} \Gamma\left(\frac{1-D}{2}\right) \sum_{n=-\infty}^{\infty} \omega_n^{2\cdot\frac{D-1}{2}}
  +C(L) \nonumber\\
  &= \frac{\tr I (\sqrt{\pi})^{D-1} }{2L^D} \Gamma\left(\frac{1-D}{2}\right)\left[ \zeta\left(1-D,\frac{\delta}{2}\right) + \zeta\left(1-D,1-\frac{\delta}{2}\right)\right]
  +C(L) , \label{EPIntSumSym}
 \end{align}
 where $\zeta(z,a)$ is the Hurwitz zeta function,
 \begin{align}
  \zeta(z,a) = \sum_{n=0}^\infty \frac{1}{(n+a)^z} .
  \label{def:zeta}
 \end{align}
The summation of the zeta functions is described by the polylogarithm, $\mathrm{Li}_s(z)$, 
 \begin{align}
  \mathrm{Li}_s(z) = \sum_{n=1}^\infty \frac{z^n}{n^s},
  \label{polylogarithm}
 \end{align}
through the Hurwitz's formula \cite{Apostol:IANT},
 \begin{align}
  \zeta\left(1-D,\frac{\delta}{2}\right) + \zeta\left(1-D,1-\frac{\delta}{2}\right)
  = \frac{4\pi }{(2\pi)^D} \frac{\Gamma\left(D\right)}{\Gamma\left(\frac{1-D}{2}\right) \Gamma\left(\frac{D+1}{2}\right)} \mathrm{Re}\,\mathrm{Li}_D\left(e^{i\pi\delta}\right) , (D > 1, 0 \leq \delta \leq 2).
 \end{align}
Therefore the effective potential $V(0)$ reduces to
 \begin{align}
  V(0)
  = \frac{\tr I}{L^D(2\sqrt{\pi})^{D-1}} \frac{ \Gamma\left(D\right)}{\Gamma\left(\frac{D+1}{2}\right) } \mathrm{Re}\,\mathrm{Li}_D\left(e^{i\pi\delta}\right)
  +C(L)  . \label{EPIntSumSym2}
\end{align}
This expression shows that the finite size effect modifies the effective potential at $\sigma=0$.

\end{document}